\documentclass[usegraphicx,usenatbib,useAMS]{mn2e} 

\def\mnras{{Mon.~ Not.~ R.~ Astron.~ Soc.~}}

\def\prd{{Phys.~ Rev.~ D.~}}

\def\apj{{Astrophys.~ J.~}}
\def\apjs{{Astrophys.~ J.~ Suppl.~}}

\def\mnras{{MNRAS}}

\def\prd{{PRD}}

\def\apj{{ApJ}}
\def\apjs{{ApJS}}

\def\aap{{A\&A}}

\newcommand{\be}{\begin{equation}}
\newcommand{\ee}{\end{equation}}
\newcommand{\ba}{\begin{eqnarray}}
\newcommand{\ea}{\end{eqnarray}}

\def\pp1{{\prime}}
\def\pp2{{\prime\prime}}

\def\2D{{\rm 2D}}

\def\Vu{{V_{\mu}}}

\def\expn{{\left<N_i\right>}}

\def\SN{{\mathcal S}/{\mathcal N}}

\def\h{{\rm h}}

\def\br{{\bf r}}

\def\bk{{\bf k}}

\def\1Loop{{\rm 1Loop}}

\def\rhob{\bar{\rho}}
\def\Hub{{\rm km}s^{-1}{\rm Mpc}^{-1}}
\def\Msol{h^{-1}M_{\odot}}

\def\Mpc{\, h^{-1}{\rm Mpc}}
\def\Mpccube{\, h^{-3} \, {\rm Mpc}^3}

\def\Gpccube{\, h^{-3} \, {\rm Gpc}^3}
\def\kMpc{\, h \, {\rm Mpc}^{-1}}

\def\dr{d^3\!r}

\def\dk{d^3\!k}

\def\dy{d^3\!y}

\def\nbar{\bar{n}}

\font\BF=cmmib10

\def\v{{\hbox{\BF v}}}

\def\by{{\hbox{\BF y}}}

\def\fun#1#2{\lower3.6pt\vbox{\baselineskip0pt\lineskip.9pt
        \ialign{$\mathsurround=0pt#1\hfill##\hfil$\crcr#2\crcr\sim\crcr}}}

\title[Covariance of cross-correlations] {Covariance of
  cross-correlations: towards efficient measures for large-scale
  structure}

\author[{\it R.~E.~Smith}]{Robert~E.~Smith$^{1}$\thanks{res@physik.unizh.ch}\\
     {$^1$ Institute for Theoretical Physics, University of Zurich, Zurich CH 8037
    }
  }

\begin{document}

\maketitle


\begin{abstract}
We study the covariance of the cross-power spectrum of different
tracers for the large-scale structure.  We develop the counts-in-cells
framework for the multi-tracer approach, and use this to derive
expressions for the full non-Gaussian covariance matrix. We show, that
for the usual auto-power statistic, besides the off-diagonal
covariance generated through gravitational mode-coupling, the
discreteness of the tracers and their associated sampling distribution
{\em can} generate strong off-diagonal covariance, and that this
becomes the dominant source of covariance as $k\gg k_f=2\pi/L$. On
comparison with the derived expressions for the cross-power
covariance, we show that the off-diagonal terms can be suppressed, if
one cross-correlates a high tracer-density sample with a low one.
Taking the {\em effective} estimator efficiency to be proportional to
the signal-to-noise ratio ($\SN$), we show that, to probe clustering
as a function of physical properties of the sample, i.e. cluster mass
or galaxy luminosity, then the cross-power approach can out perform
the auto-power one by factors of a few.  We confront the theory with
measurements of the mass-mass, halo-mass, and halo-halo power spectra
from a large ensemble of $N$--body simulations. We show that there is
a significant $\SN$ advantage to be gained from using the cross-power
approach when studying the bias of rare haloes. The analysis is
repeated in configuration space and again $\SN$ improvement is found.
We estimate the covariance matrix for these samples, and find strong
off-diagonal contributions. The covariance depends on halo mass, with
higher mass samples having stronger covariance. In agreement with
theory, we show that the covariance is suppressed for the
cross-power. This work points the way towards improved estimators for
studying the clustering of tracers as a function of their physical
properties.
\end{abstract}



\section{Introduction}

The power spectrum of matter fluctuations is of prime concern in
cosmology, since it contains detailed information about the underlying
world model and provides a method for probing the initial conditions
of the Universe. Moreover, if the statistical properties of the
initial fluctuations form a Gaussian Random Field, as is the case for
most inflationary models, then the power spectrum provides a complete
description for the spatial statistics of the density
field. Consequently, over the last few decades a large fraction of
observational and theoretical effort has been invested in estimating
the power spectrum of galaxies from large redshift surveys and also to
devising methods for extracting cosmological information from the
signal
\citep{Feldmanetal1994,PeacockDodds1994,Percivaletal2001,Tegmarketal2004a,Coleetal2005,Tegmarketal2006,Percivaletal2007a}.

In order to obtain robust cosmological constraints from such data
sets, one, however, requires additional knowledge about the signal
covariance matrix -- or the correlation function of power
fluctuations.  Unlike the power spectrum, which is the Fourier
transform the two-point correlation function, the covariance has had
relatively little attention. This mainly stems from the fact that in
order to estimate this quantity from a galaxy survey, or to compute it
theoretically, one is required to investigate the four-point function
of Fourier modes, more commonly the trispectrum of galaxies, and this
is a substantially more complex quantity.

The first study of power spectrum covariance, in the modern context,
was performed by \citet{Feldmanetal1994}, who showed, under the
assumption of Gaussianity, that the matrix was diagonal and that the
variance per band-power was proportional to the square of the power in
the band \citep[see also][]{StirlingPeacock1996}. This result gave
impetus to those advocating the use of power spectra for large-scale
structure work, over the simpler two-point correlation function,
$\xi$, since under these same assumptions $\xi$ possesses correlated
errors \citep{Bernstein1994}.

Later, both \citet{Scoccimarroetal1999b} and \citet{MeiksinWhite1999}
independently showed that the real situation was much more complicated
than the Gaussian calculation would lead one to believe.  They
recognized that nonlinear gravitational instability would cause
different Fourier modes to become coupled together, thus breaking the
Gaussianity. In \citet{Scoccimarroetal1999b}, this mode-coupling
behaviour was demonstrated by using higher-order perturbation theory
to calculate the trispectrum and by an analysis of results from an
ensemble of $N$-body simulations. One direct consequence of this
effect, was that the fractional errors on the dark matter power
spectrum were shown to reach an almost constant plateau on
intermediate to small scales, regardless of the additional number of
Fourier modes \citep[see
  also][]{ScoccimarroSheth2002,Hamiltonetal2006,RimesHamilton2006,Takahashietal2009}. They
also showed that off-diagonal covariance on small scales was
generated, but their results on large scales appeared inconclusive,
owing to small volumes and hence increased sample variance.

\citet{MeiksinWhite1999} reached similar conclusions. They also
extended the theoretical analysis to include the covariance in the
power spectrum, arising from the finite sampling of the density field,
referred to as Poisson sampling variance. It is well known that this
is of importance for correctly determining the diagonal errors of the
covariance matrix for rare tracers of the density field, such as
bright galaxies and clusters.  Whilst the covariance matrix of the
dark matter power spectrum has been studied in some detail, that of
haloes and galaxies has not received nearly the same level of
attention -- at least not beyond the assumption of linear density
evolution and linear biasing. Notable contributions are:
\citet{CoorayHu2001,ScoccimarroSheth2002,Sefusattietal2006,Anguloetal2008a}.
However, as we will show for the first time in this work, the
discreteness terms that were neglected by \citet[since they were
  mainly studying the dark matter clustering][]{MeiksinWhite1999},
inevitably, become the dominant source of off-diagonal error for
discrete tracers of the mass distribution. 

Recently, cross-correlation techniques have become an ever more
important tool for extracting information from large-scale structure
data. For instance, in a recent theoretical study,
\citet{Smithetal2007} demonstrated, using $N$-body simulations, that
the cross-power spectrum between dark matter and haloes had several
advantages over the simpler auto-power spectrum method. In particular,
a reduced shot-noise correction and noise properties. This
cross-correlation approach has been further exploited to elucidate the
environmental dependence of halo bias
\citep{Jingetal2007,Anguloetal2008b} and recently as a means for
probing the large-scale scale dependence of bias in models of
primordial Non-Gaussianity
\citep{Dalaletal2008,Desjacquesetal2009,Pillepichetal2008,Grossietal2009}. Also,
the cross-correlation approach has recently been applied to real
survey data, to study the intrinsic clustering properties of quasars
in the SDSS photometric redshift catalogue, through cross-correlating
them with the more abundant Luminous Red Galaxy (LRG) sample
\citep{Padmanabhanetal2008b}. It is therefore of great use to have an
explicit calculation for the covariance of cross-correlations for use
in likelihood analysis.  Moreover, the covariance matrix is an
important ingredient for any Fisher matrix parameter forecast, and
hence an essential tool for optimal survey design \citep{Tegmark1997}.

The paper is broken up as follows: In \S\ref{sec:background} we
develop the standard counts-in-cells framework to calculate the
cross-power spectrum of two different tracers of the large scale
structure. In the analysis, we pay close attention to the assumed
sampling distribution: besides the usual Poisson model, we also
consider the toy-model scenario where one tracer is simply a
sub-sample of the other and results are presented for both cases.
This is instructive, since it is likely that not all galaxies are
equally good tracers of the mass -- in particular those hosted in the
same halo. Then in \S\ref{sec:crosscovar} we derive an expression for
the covariance of the cross-power spectrum, including all non-Gaussian
and finite sampling contributions to the error. Limiting cases are
considered and expressions are also given for band-power averages.  We
evaluate the expected covariance signal for several different tracers
of the mass. In \S\ref{sec:Efficiency} we compare the efficiency of
the cross-power approach with that of the simpler auto-power approach.
In \S\ref{sec:CovCorr} the analogous expressions are derived for the
cross-correlation function. In \S\ref{sec:simulations} we make a
direct comparison of the theoretical predictions with estimates
measured from the {\tt zHORIZON} simulations, a large ensemble of
dark matter $N$-body simulations with total volume
$\sim100\Gpccube$. Finally in \S\ref{sec:conclusions} we summarize our
results and conclude.


\section{Counts-in-cells framework for multiple tracers}\label{sec:background}

\subsection{Statistics of a single tracer population}\label{ssec:onetracer}

Consider a single population of $N$ discrete objects in some large
volume $V_{\mu}$ that trace the large-scale structure of the Universe
in some way. Following \citet{Peebles1980}, we shall assume that these
tracers are Poisson sampled from some underlying smooth density field,
and that the statistics of this underlying field are well described by
a Gaussian Random Field. Hence, on partitioning space into a set of
infinitesimal volume elements $\delta V$, the probability of finding
$N_i$ galaxies in an element at position vector $\br_i$ is given by
\ba P(N_i|\lambda =  n(\br_i)\delta V) & = & \frac{\exp[-\lambda]\lambda^{N_i}}{N_i!} 
\nonumber \\
& \approx & \left\{
\begin{array}{ll}
n(\br_i)\delta V & (N_i=1)    \\ 
1-n(\br_i) \delta V & (N_i=0) \\
0 & (N_i>1)
\end{array}
\right. , \ea
where $n(\br)$ is the continuous number density function for tracers
in the volume, which, in the local model for galaxy bias
\citep{FryGaztanaga1993,Coles1993}, is directly related to the
underlying distribution of fluctuations in the CDM; and for the
linearized relation this is simply:
$n(\br)=\nbar\left[1+b\delta(\br)\right]$, where
$\delta(\br)=\left[\rho(\br)-\rhob\right]/\rhob$ is the fractional
over-density in the dark matter relative to the mean density
$\rhob$. The probabilities of finding $N_i\ge2$ are higher powers of
the infinitesimal quantity $\delta V$ and so are negligible. It can
now be shown that all of the one-point moments are $(m\ge1)$:
\be \left<N_i^m\right>_{p,s} = \dots = \left<N_i\right>_{p,s} 
= \left<n(\br_i)\delta V\right>_s = \nbar\delta V\ ; \ee
and the central moments of the distribution are $(m>1)$
\be \left<(N_i-\expn)^m\right>_{p,s} = \left<n(\br_i)\delta V\right>_s = 
\nbar\delta V\  \ ,\ee
where in the above we used the notation $\left<\dots\right>_{p,s}$ to
denote an averaging over all possible samplings of the points $p$ and
all points in space $s$ (for brevity we shall simply write
$\left<\dots\right>$).

The two-point moments may also be derived. Consider the joint
probability of finding objects in two disjoint volume elements $\delta
V_i$ and $\delta V_j$ separated by a vector $\br_{ij}=\br_i-\br_j$, in
the Poisson sampling model this is given simply by the product of the
independent probabilities $(i\ne j)$:
\ba 
P(N_i,N_j) & = & P(N_i)P(N_j)\  \\
& = & n(\br_i)n(\br_j)\delta V_i \delta V_j \ .
\label{eq:2pointprob}\ea
On averaging the two-point moments may be written,
\be \left<N^k_i N^m_j\right> = 
\nbar^2\delta V_i\delta V_j [1+\xi({\br_i},\br_j)] \ ,\ee
where $\nbar\equiv\left<n(\br)\right>=\sum_i N_i/\Vu=N/\Vu$ is the
mean number density of tracers and $\xi({\br_i},\br_j)$ is the
two-point auto-correlation function. Hence, correlations are
introduced into the sample, if and only if the points in the
underlying continuous field are correlated.


\subsection{The auto-power spectrum}\label{ssec:autopower}

We define the Fourier relations for the density field as,
\ba 
\delta(\br) & = & \frac{\Vu}{(2\pi)^3}\int \dk \,
\delta(\bk) \exp(-i\bk\cdot\br) \ ; \\
\delta(\bk) & = & \frac{1}{\Vu}\int \dr\, \delta(\br) \exp(i\bk\cdot\br)\ .
\ea
The density field of the discrete counts in cells is written
\be \delta^d(\br)=\frac{1}{\nbar}\sum_i
\left[N_i-\left<N_i\right>\right]\delta^D(\br-\br_i)\ , \ee
which on insertion into our definition of the Fourier transform leads to 
the discrete sum
\be \delta^d(\bk)=\frac{1}{N}\sum_i \left[N_i-\left<N_i\right>\right]
 \exp(i\bk\cdot\br_i) \ .\ee

We may now compute the power spectrum of the discrete set of tracers,
\[
\left<\delta^d(\bk_1)\delta^d(\bk_2)\right>  =   
\frac{1}{N^2}\sum_{i,j} \left<\left[N_i-\left<N_i\right>\right]
\left[N_j-\left<N_j\right>\right]\right> \]
\be \hspace{2.4cm} \times e^{i\bk_1\cdot\br_i+i\bk_2\cdot\br_j} \ \ee
\[ \hspace{2.4cm}= \frac{1}{\Vu^2}\sum_{i\ne j} 
\delta V_i\delta V_j \xi(\br_i,\br_j) 
e^{i\bk_1\cdot\br_i+i\bk_1\cdot\br_j}\]  
\be \hspace{2.4cm} +  \frac{1}{N\Vu}\sum_{i=j} 
\delta V_i e^{i(\bk_1+\bk_2)\cdot\br_i}\ .
\ee
The sums over cells can be transformed into volume integrals, and the
double volume integral over the correlation function in the first term
can be simplified by recalling that through statistical homogeneity
$\xi(\br_i,\br_j)=\xi(\br_i-\br_j,0)$. We may then make use of the
orthogonality of the Fourier basis functions to evaluate sums of the
type,
\be \sum_i \delta V_i e^{i(\bk_1+\bk_2)\cdot \br_i} = \Vu \delta^K_{\bk_1,-\bk_2} \ .\ee
Hence, after performing these steps and introducing our definition of
the power spectrum as 
\be P(\bk_1)\delta^{K}_{\bk_1,-\bk_2} \equiv \Vu
\left<\delta(\bk_1)\delta(\bk_2)\right>\ ,\ee
we recover the standard result for the power spectrum of discrete
tracers \citep{Peebles1980}:
\be P^d(k)=P^c(k)+\frac{1}{\nbar} \ , \label{eq:autospectrum}\ee
where $P^c$ is the power spectrum of the underlying continuous field
of tracers. The constant term on the right-hand-side of the equation
is more commonly referred to as the `shot-noise correction' term, and
is the additional variance introduced through discreteness.


\subsection{Statistics of two tracer populations}\label{ssec:twotracers}

We shall now extend the above formalism to the problem of two
different tracer populations, which we shall denote as $A$ and
$B$. Let the total number of objects in samples A and B be $N_A$ and
$N_B$, and the numbers of each type of object in the $i$th cell be
$N_{A,i}\equiv N_{A}(\br_i)$ and $N_{B,i}\equiv N_{B}(\br_i)$,
respectively. Likewise the mean number densities are $\nbar_A$ and
$\nbar_B$. We now consider two cases for the sampling distributions,
these are:

\begin{enumerate}
\item {\bf Non-overlapping tracers}. $A$ and $B$ are both independent
  Poisson samples of the underlying continuous density field. In this
  case the joint probability distribution for obtaining objects of
  types $A$ and $B$ in a single cell is:
\[ P(N_{A,i},N_{B,i}) = P(N_{A,i}|\lambda_A)P(N_{B,i}|\lambda_B)\ ;
\]
\be  
\approx \left\{
\begin{array}{ll}
1-\left[n_A(\br)+n_B(\br)\right]\delta V  & (N_A=0,N_B=0) \\
n_A(\br)\delta V              & (N_A=1,N_B=0) \\
n_B(\br)\delta V              & (N_A=0,N_B=1) \\
0                            & (N_A\ge1,N_B\ge1) \ .
\end{array}
\right.
\ee
The one-point cross-moments are then calculable $(m\ge1, k\ge1)$,
\be \left<N^m_{A,i} N^k_{B,i}\right>=0 \ ;\ee
and so too the central moments: 
\be \left< \left(N_{A,i} -\left<N_{A,i}\right>\right)^m
\left(N_{B,i}-\left<N_{B,i}\right>\right)^k \right>=0 \ .\ee
As in Eq.~(\ref{eq:2pointprob}), the two-point cross-moments may also
be derived and these are $(i\ne j):$
\be 
\left<N^m_{A,i} N^k_{B,j}\right>=\nbar_A\nbar_B \delta V_i\delta V_j
[1+\xi^{AB}(\br_i,\br_j)] \ ,\ 
\ee
where $\xi^{AB}$ is the two-point cross-correlation function of the
tracers A and B.
\vspace{3mm}
\item {\bf Overlapping tracers}. $A$ is a Poisson sample of the
underlying continuous density field, and $B$ is a sub-sample of
$A$. This time the joint probability distribution for obtaining
objects of types $A$ and $B$ is written:
\be P(N_{A,i},N_{B,i}) = P(N_{A,i}|\lambda_A)P(N_{B,i}|N_{A,i})\ .\ee 
The conditional probability $P(N_{B,i}|N_{A,i})$ is the key object of
interest here, and as a simple illustrative example we will take this
as:
\be P(N_{B,i}|N_{A,i})=
\left\{
\begin{array}{ll}
1   &  (N_B=0|N_A=0) \\
a   &  (N_B=1|N_A=1) \\
1-a &  (N_B=0|N_A=1) \\
0   &  (N_B>1|N_A\ge1)\ , 
\end{array}
\right.
\ee 
where we shall fix $a\equiv N_B/N_A$. Again, the one-point
cross-moments are also calculable for this sampling model,
\be \left<N^m_{A,i} N^k_{B,i}\right>= a\,\bar{n}_A\delta V = \bar{n}_B\delta V\ ;\ee
and so too the central moments,
\be \left<
\left(N_{A,i} -\left<N_{A,i}\right>\right)^m
\left(N_{B,i} -\left<N_{B,i}\right>\right)^k
\right>= \bar{n}_B\delta V\ .\ee
Similarly, the two-point cross-moments are also calculable,
\be 
\left<N^m_{A,i}N^k_{B,j}\right>=\nbar_A\nbar_B \delta V_i\delta V_j
[1+\xi^{AB}(\br_i,\br_j)] \ .\ 
\ee
\end{enumerate}


\subsection{The cross-power spectrum}\label{ssec:crosspower}

We may also compute the cross-power spectrum of tracers A and B, and
for both the non-overlapping (case i) and overlapping (case ii)
sampling distributions. The Fourier modes for tracers A and B can be
written,
\be \delta^d_{T}(k)=\frac{1}{N_{\rm T}}
\sum_i \left[N_{\rm T}(\br_i)-\left<N_{{\rm T},i}\right>\right]
 \exp(i\bk\cdot\br_i) \ ,\ee
where ${\rm T}=\{A,B\}$ denotes the tracer type. As for the
auto-spectrum, we shall define the cross-power spectrum, as
\be P_{AB}(\bk_1)\delta^{K}_{\bk_1,-\bk_2}\equiv
\Vu\left<\delta_A(\bk_1) \delta_B(\bk_2)\right> \ . \ee
Following now the steps in \S\ref{ssec:autopower}, but this time using
the statistics for the counts-in-cells as given in the previous
section, we find that the cross-power of discrete tracers A and B
obeys the relation:
\be P^d_{AB}(k)=P^{c}_{AB}(k)+\left\{\frac{1}{\nbar_A}\right\} \
\label{eq:crosspower} ,\ee
where $P^{c}_{AB}$ is the cross-power spectrum of the underlying
continuous fields. This expression is almost identical to the result
for the auto-spectrum (Eq.~\ref{eq:autospectrum}), however we
emphasize an important difference, the constant term is enclosed by
curly brackets. {\em In this paper $\{\dots\}$ shall have the special
  meaning that this term only appears when there is an overlap between
  samples A and B, as in sampling case(ii), otherwise this term is
  exactly zero \citep[see][]{Peebles1980}}.  We note that this
notation shall be exploited throughout the rest of the paper, to
represent the results from both sampling distributions with a single
equation. More intuitively, the appearance of the constant term in the
cross-power spectrum warns us that, if the two samples are not truly
independent, then we should expect some finite sampling correction.


\section{Covariance of the cross-power spectrum}\label{sec:crosscovar}

We now turn to the calculation of the full non-Gaussian covariance of
the cross-power spectrum for discrete tracers A and B.  Note that,
when considering sampling case ii, and in the limit that $N_A=N_B$,
then we shall recover the standard covariance relations for the
auto-power spectrum \citep{Scoccimarroetal1999b,MeiksinWhite1999}.


\subsection{Definition of the covariance}

To begin, we define the covariance, per mode, of the cross-power
spectrum for discrete tracers A and B as,
\ba C_{AB}^{d} & \equiv & {\rm
  Cov}\left[P^{d}_{AB}(\bk_1),P^{d}_{AB}(\bk_2)\right] \nonumber \\ &
= & \left<P^{d}_{AB}(\bk_1)P^{d}_{AB}(\bk_2)\right>-
\left<P^{d}_{AB}(\bk_1)\right>\left<P^{d}_{AB}(\bk_2)\right> \ .\ea
On inserting the definition for the cross-power spectrum,
$P_{AB}\equiv \Vu \left<\delta_{A}(\bk_1)\delta_{B}(-\bk_1)\right>$,
and making use of Eq.~(\ref{eq:crosspower}) in the second term on the
right-hand-side, then we obtain
\ba 
C_{AB}^{d} & = &
\Vu^2 \left<\delta^d_A(\bk_1)\delta^d_B(-\bk_1)\delta^d_A(\bk_2)\delta^d_B(-\bk_2)\right> \\
& & -
\left(P^c_{AB}(\bk_1)+\left\{\frac{1}{\nbar_A}\right\}\right)
\left(P^c_{AB}(\bk_2)+\left\{\frac{1}{\nbar_A}\right\}\right)\ .
\ea
Thus we see that in order to compute the covariance of the cross-power
spectrum, it is also necessary to evaluate the four-point function of
Fourier modes, or more commonly the trispectrum.


\subsection{Evaluating the discrete cross-trispectrum}

Using the counts-in-cells approach, the four point cross-correlation
function of Fourier modes can be written explicitly as,
\[ \left<\delta^d_A(\bk_1)\delta^d_A(\bk_2)\delta^d_B(\bk_3)\delta^d_B(\bk_4)\right>
=\frac{1}{N_A^2}\frac{1}{N_B^2}\sum_{i,j,k,l} e^{i\bk_1\cdot\br_i+\dots+i\bk_4\cdot\br_l}
\]
\[  \hspace{1cm}\times 
\left< \frac{}{}
\left(N_{A,i}-\left<N_{A,i}\right>\right)
\left(N_{A,j}-\left<N_{A,j}\right>\right)\right. 
\]
\be \hspace{1cm}\times 
\left. \frac{}{}
\left(N_{B,k}-\left<N_{B,k}\right>\right)
\left(N_{B,l}-\left<N_{B,l}\right>\right)
\right> \ \label{eq:fourpoint1}\ .\ee
Thus we find that in order to evaluate the trispectrum, we are in turn
required to evaluate the four-point cross-correlation function of
counts-in-cells. Again, following \citet{Peebles1980}, we break this
quadruple sum into five types of terms, each of which arises from a
particular partitioning of the indices $(i,j,k,l)$. Full details are
presented in the following subsections, those not wishing to be
embattled at this stage should skip ahead to \S\ref{ssec:FullNGCov}.


\subsubsection{Terms $(i \ne j \ne k \ne l)$} 

Terms in the sum with these indices, correspond to contributions to
the product from the connected four-point correlation function of the
field. These terms can be rewritten as,
\[ \left<\left(N_{A,i}-\left<N_{A,i}\right>\right)\dots
\left(N_{B,l}-\left<N_{B,l}\right>\right)\right>
=
\nbar_A^2\nbar_B^2 \delta V_i\dots \delta V_l\]
\be \hspace{1cm} \times
\left[ \eta_{ijkl}^{AABB}+
\xi^{AA}_{ij}\xi^{BB}_{kl}+
\xi^{AB}_{ik}\xi^{AB}_{jl}+
\xi^{AB}_{il}\xi^{AB}_{jk}\right]
\ ,\ee
where for convenience we have introduced the abbreviated notation for
the two-, three- and four-point correlation functions:
\[
\xi_{ij} \equiv \xi(\br_i,\br_j) \ ; \ 
\zeta_{ijk}  \equiv \zeta(\br_i,\br_j,\br_k) \ ; \
\eta_{ijkl} \equiv \eta(\br_i,\br_j,\br_k,\br_l) \ .
\]
On inserting the above expression into Eq.~(\ref{eq:fourpoint1}),
transforming the sums over cells to volume integrals, and using the
statistical homogeneity of the correlation functions, we obtain the
following expression,
\[ 
\Vu^2 \left<\delta^d_A(\bk_1)\dots\delta^d_B(\bk_4)\right>
=\frac{1}{\Vu}T_{AABB}(\bk_1,\bk_2,\bk_3,\bk_4)\delta^{K}_{\bk_1+\dots+\bk_4,0} 
\]
\[ \hspace{1cm} + 
 P_{AA}(\bk_1)P_{BB}(\bk_3)
\delta^{K}_{\bk_1,-\bk_2}\delta^{K}_{\bk_3,-\bk_4}
\]
\[ \hspace{1cm} + 
P_{AB}(\bk_1)P_{AB}(\bk_2)
\delta^{K}_{\bk_1,-\bk_3}\delta^{K}_{\bk_2,-\bk_4}
\]
\be \hspace{1cm} + 
P_{AB}(\bk_1)P_{AB}(\bk_2)
\delta^{K}_{\bk_1,-\bk_4}\delta^{K}_{\bk_2,-\bk_3}\ ,
\label{eq:tri1} \ee
where the irreducible or connected trispectrum of the underlying
continuous density field has been defined as,
$T(\bk_1,\dots,\bk_4)\equiv \Vu^3
\left<\delta(\bk_1)\dots\delta(\bk_4)\right>_c
\delta^K_{\bk_1+\dots\bk_4,0}$. This obeys a Fourier relation with the
irreducible four-point correlation function $\eta_{ijkl}$.


\subsubsection{Terms $(i\ne j\ne k = l)$ + perms.}

There are six types of term that arise from the equivalence of two of
the indices, and in order to evaluate these, we are required to deal
with products of the form,
\[ \left<
\left(N_{A,i}-\left<N_{A,i}\right>\right)
\left(N_{A,j}-\left<N_{A,j}\right>\right)
\left(N_{B,k}-\left<N_{B,k}\right>\right)^2 
\right> 
\]
\be
=
\left<
\left(N_{A,i}-\left<N_{A,i}\right>\right)
\left(N_{A,j}-\left<N_{A,j}\right>\right)
N_{B,k}\right> \ ,
\ee
\be
=\nbar_{A}^2\nbar_B\delta V_i\delta V_j\delta V_j
\left[\xi_{ij}^{AA}+\zeta_{ijk}^{AAB}\right] \ ,
\ee
where the second equivalence follows from the rules for the
cross-moments \S\ref{ssec:twotracers}. Hence, on repeating this
procedure for all possible ways of equivalencing two indices we
arrive at six expressions. Then, on following a procedure similar to
the evaluation of the cross-power spectrum, and on introducing the
bispectrum $B$ as,
\be B(\bk_1,\bk_2)\delta^{K}_{\bk_1+\bk_2+\bk_3,0}\equiv \Vu^2
\left<\delta(\bk_1)\delta(\bk_2)\delta(\bk_3)\right> \ , \ee 
and noting that $B$ and $\zeta$ are Fourier duals, we find that these
terms can be written:
\[ 
\Vu^2 \left<\delta^d_A(\bk_1)\dots\delta^d_B(\bk_4)\right>=
\]
\[  
\left[
\frac{1}{\nbar_B}P_{AA}(\bk_1)+\frac{1}{\nbar_A}P_{BB}(\bk_3)\right]
\delta^{K}_{\bk_1,-\bk_2}
\delta^{K}_{\bk_3,-\bk_4}+
\]
\[
\left\{\frac{1}{\nbar_A}\left[P_{AB}(\bk_1)+P_{AB}(\bk_2)\right]
\left[
\delta^{K}_{\bk_1,-\bk_4}\delta^{K}_{\bk_2,-\bk_3}
+\delta^{K}_{\bk_1,-\bk_3}\delta^{K}_{\bk_2,-\bk_4}
\right]\right\}
\]
\[ 
+ \left[\frac{1}{N_B}B_{AAB}(\bk_1,\bk_2) + \frac{1}{N_A}B_{ABB}(\bk_3,\bk_4)\right]
\delta^{K}_{\bk_1+\dots+\bk_4,0}
\]
\[
+ \left\{\frac{1}{N_A} 
\left[
B_{ABB}(\bk_1,\bk_3) +B_{ABB}(\bk_1,\bk_4)+B_{ABB}(\bk_2,\bk_3) \right.\right.
\]
\be 
\left. \left.  +B_{ABB}(\bk_2,\bk_4) \right]
\delta^{K}_{\bk_1+\dots+\bk_4,0}\frac{}{} \right\}
\label{eq:tri2}  \ ,
\ee
where $B_{ABB}$ and $B_{AAB}$ are the cross-bispectra of the fields A
and B. 


\subsubsection{Terms $(i=j\ne k = l)$ + perms.}

There are three terms of this form that arise in the quadruple sum,
and theses involve evaluation of quantities of the form:
\[ \left<
\left(N_{A,i}-\left<N_{A,i}\right>\right)^2
\left(N_{B,k}-\left<N_{B,k}\right>\right)^2 
\right> = \left<N_{A,i}N_{B,k}\right> 
\]
\be
\hspace{1cm}=\nbar_{A}\nbar_B\delta V_i\delta V_k
\left[1+\xi_{ik}^{AB}\right] \ ,
\ee
where again we have used the relations for the cross-moments from
\S\ref{ssec:twotracers}. On repeating this procedure for the other two
terms, and repeating the analysis as before, we find that these types
of terms can be written together as,
\[ 
\Vu^2 \left<\delta^d_A(\bk_1)\dots\delta^d_B(\bk_4)\right>=
\frac{\delta^{K}_{\bk_1,-\bk_2}\delta^{K}_{\bk_3,-\bk_4}}{\nbar_A\nbar_B}
\]
\[ 
+\left\{\frac{1}{\nbar_A^2}
\left[\frac{}{}
\delta^{K}_{\bk_1,-\bk_3}\delta^{K}_{\bk_2,-\bk_4}+
\delta^{K}_{\bk_1,-\bk_4}\delta^{K}_{\bk_2,-\bk_3}
\right]\right\}
\]
\[
+\frac{1}{\nbar_A\nbar_B\Vu}P_{AB}(\bk_1+\bk_2)\delta^{K}_{\bk_1+\dots+\bk_4,0} 
\]
\be + 
\left\{
\frac{1}{\nbar_A^2\Vu}\left[P_{BB}(\bk_1+\bk_3)+P_{BB}(\bk_1+\bk_4)\right]
\delta^{K}_{\bk_1+\dots+\bk_4,0}\right\}
\label{eq:tri3}\ .
\ee


\subsubsection{Terms $(i = j = k \ne l)$ + perms.}

There are four possible types of term that arise from this combination
of indices and each of these requires us to evaluate a product like:
\[ \left<
\left(N_{A,i}-\left<N_{A,i}\right>\right)^2
\left(N_{B,i}-\left<N_{B,i}\right>\right)
\left(N_{B,l}-\left<N_{B,l}\right>\right)
\right> 
\]
\[ 
\hspace{1cm}= \left< N_{B,i}\left(N_{B,l}-\left<N_{B,l}\right>\right)\right> 
\]
\be 
\hspace{1cm}=\nbar_B^2\xi_{il}^{BB}\delta V_i \delta V_l\ .
\ee
Hence, on repeating this for the four possible arrangements of the
indices, and on using the methods described for the previous terms,
we find that all of these terms reduce to the following expression:
\[
\Vu^2 \left<\delta^d_A(\bk_1)\dots\delta^d_B(\bk_4)\right>=
\left\{\frac{1}{\nbar_A\nbar_B\Vu}\left[
P_{AB}(\bk_1)+
P_{AB}(\bk_2)
\right]
\right.
\]
\be 
\hspace{1cm}\left.
+\frac{1}{\nbar_A^2\Vu}
\left[
\frac{}{}
P_{BB}(\bk_3)+
P_{BB}(\bk_4)\right]
\right\}\delta^{K}_{\bk_1+\dots+\bk_4,0}
\label{eq:tri4}\ee


\subsubsection{Terms $(i=j=k=l)$}

There is only one form for this type of term in the quadruple sum,
and to evaluate it we are required to compute the quantity,
\[ \left<
\left(N_{A,i}-\left<N_{A,i}\right>\right)^2
\left(N_{B,i}-\left<N_{B,i}\right>\right)^2 \right> =
\left<N_{B,i}\right>=\nbar_B \delta V_i \ .
\]
Hence, this has the form
\be 
\Vu^2 \left<\delta^d_A(\bk_1)\dots\delta^d_B(\bk_4)\right>
= \left\{\frac{1}{\nbar_A^2\nbar_B V_{\mu}} \right\} \delta^{K}_{\bk_1+\dots+\bk_4,0}\ .
\label{eq:tri5}\ee


\subsection{Expressions for the cross-power covariance}\label{ssec:FullNGCov}

The summation of
Eqns~(\ref{eq:tri1},\,\ref{eq:tri2},\,\ref{eq:tri3},\,\ref{eq:tri4},\,\ref{eq:tri5})
gives the complete description of all the terms entering the
cross-trispectrum of Fourier modes for samples A and B. We may now use
this to obtain the full non-Gaussian covariance of the cross-power
spectrum in two different modes $\bk_1$ and $\bk_2$.  To do this we
simply take
Eqns~(\ref{eq:tri1},\,\ref{eq:tri2},\,\ref{eq:tri3},\,\ref{eq:tri4},\,\ref{eq:tri5}),
and set the arguments of the wave modes to be
\[ (\bk_1,\bk_2,\bk_3, \bk_4)\rightarrow
(\bk_1,\bk_2,-\bk_1,-\bk_2)\ .\]
This gives us the quantity
$\left<\delta^d_A(\bk_1)\delta^d_B(-\bk_1)\delta^d_A(\bk_2)\delta^d_B(-\bk_2)\right>$.
Hence, the covariance is given by:
\[
C^d_{AB}=\frac{1}{\Vu}T_{AABB}(\bk_1,\bk_2,-\bk_1,-\bk_2) 
\]
\[
+\left( P_{AA}(\bk_1)+\frac{1}{\nbar_A}\right)\left(P_{BB}(\bk_2)+\frac{1}{\nbar_B}\right)
\delta^{K}_{\bk_1,-\bk_2}
\]
\[
+\left( P_{AB}(\bk_1)+\left\{\frac{1}{\nbar_A}\right\}\right)
 \left( P_{AB}(\bk_2)+\left\{\frac{1}{\nbar_A}\right\}\right)
\delta^{K}_{\bk_1,\bk_2}+
\]
\[
\frac{B_{AAB}(\bk_1,\bk_2)}{N_B}+\frac{1}{N_A} \left[\frac{}{} 
B_{ABB}(-\bk_1,-\bk_2)+\left\{\frac{}{} B_{ABB}(\bk_1,-\bk_2)\right.\right.
\]
\[ \left.\left. +\frac{}{}B_{ABB}(\bk_2,-\bk_2)
+B_{ABB}(\bk_2,-\bk_1)+B_{ABB}(\bk_1,-\bk_1)
\right\}\right]
\]
\[
+\frac{P_{AB}(\bk_1+\bk_2)}{\nbar_A\nbar_B \Vu}
+ \left\{ 
\frac{1}{\nbar_A^2\Vu}\left[P_{BB}({\bf 0})+P_{BB}(\bk_1-\bk_2)\right]
\right\}
\]
\[
+\left\{
\frac{1}{\nbar_A^2\Vu}\left[P_{BB}(-\bk_1)+P_{BB}(-\bk_2)\right]
\right\}
\]
\be
+\left\{
\frac{1}{\nbar_A\nbar_B\Vu}\left[P_{AB}(\bk_1)+P_{AB}(\bk_2)\right]\right\}+
\left\{\frac{1}{\nbar_A^2\nbar_B V_{\mu}} 
\right\} \label{eq:CovCrossPowFull}\ .
\ee
Again, we remind the reader that the terms in curly brackets vanish
for the case where samples A and B have no overlap. It should also be
noted that when samples A and B are equivalent, then we recover the
expressions for the covariance of the auto-power spectrum
\citep{Scoccimarroetal1999b,MeiksinWhite1999}.


\begin{figure*}
\centerline{ 
\includegraphics[width=8cm]{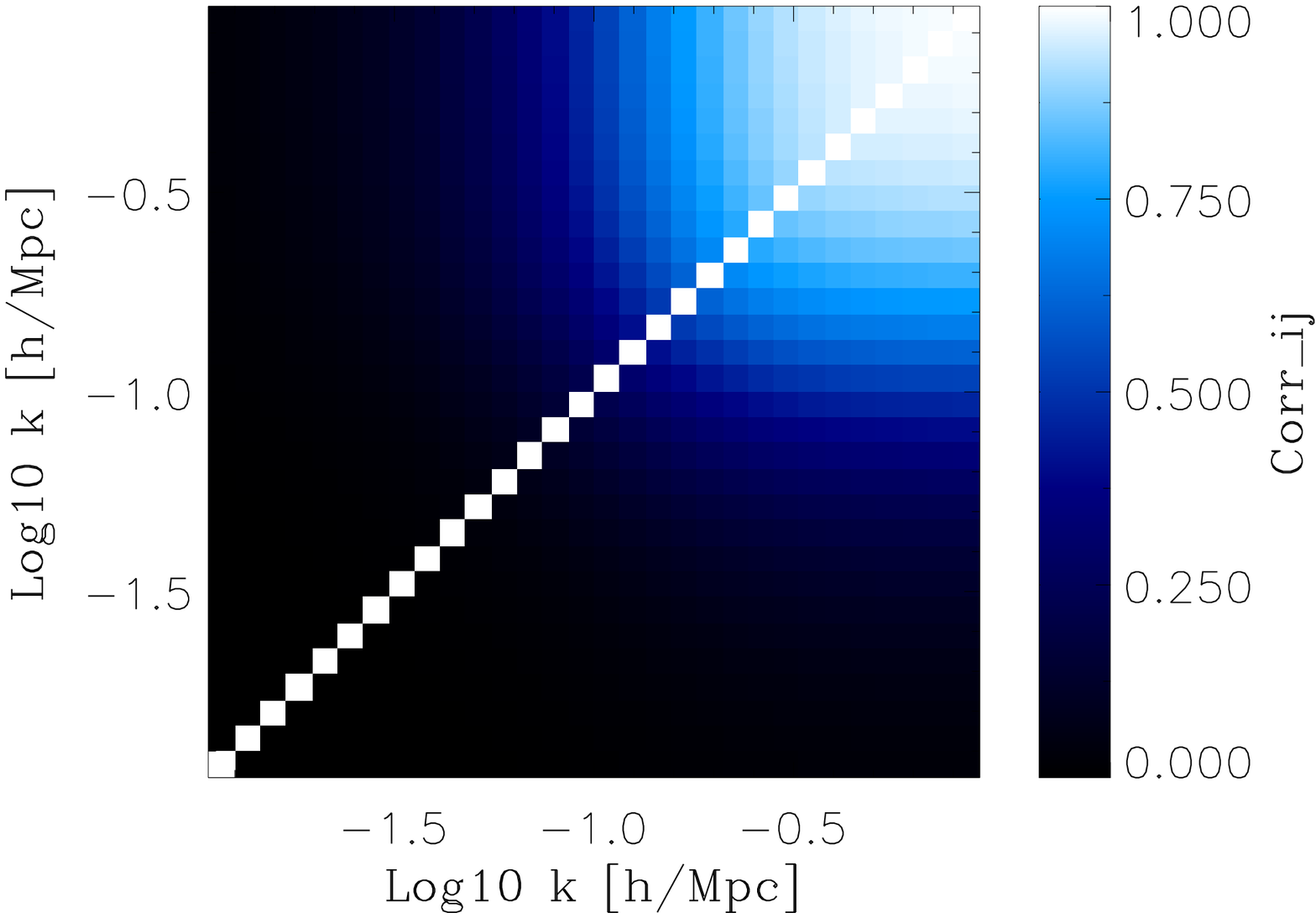}\hspace{1cm}
\includegraphics[width=8cm]{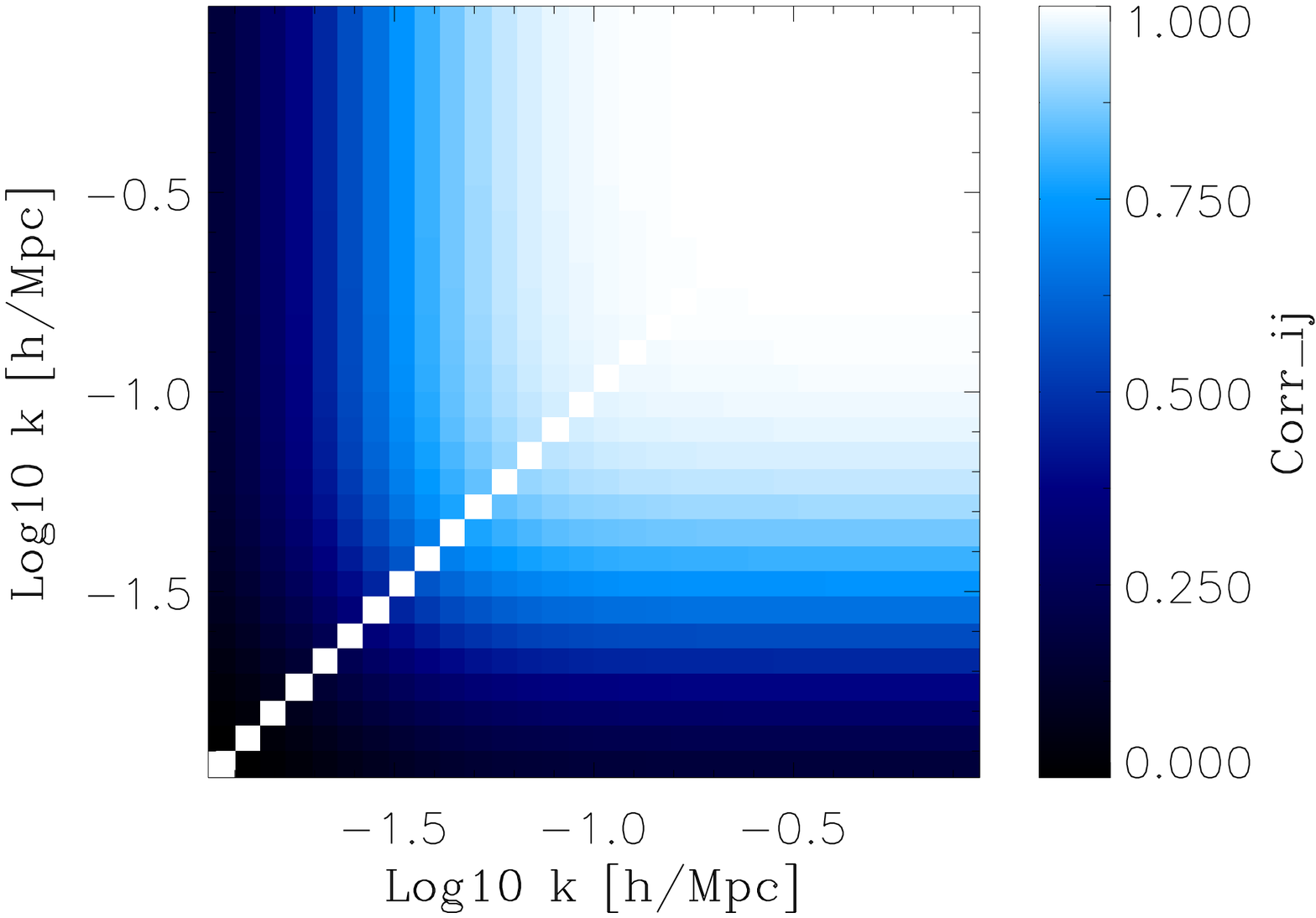}}
\caption{\small{Theoretical predictions for the halo-halo auto-power
    spectrum correlation matrix as a function of wavenumbers $k_i$ and
    $k_j$. Here all of the covariance is generated by the discreteness
    terms and all non-Gaussian terms generated through gravitational
    instability have been set to zero, i.e. $T_c=B_c=0$. {\em Left
      panel}: dark matter haloes with masses
    $M\in[1.0\times10^{13},2.0\times10^{13}]\Msol$. {\em Right panel}:
    dark matter haloes with masses
    $M\in[1.0\times10^{15},2.0\times10^{15}]\Msol$.}
\label{fig:CorrMatrixHH}}
\end{figure*}


\subsection{Band-power average covariance}

The above formula gives us the full expression for the covariance in
the cross-power spectrum per Fourier mode. In practice, the power is
estimated by averaging over all wavemodes in thin spherical shells in
$k$-space -- band-powers. The band-power average power spectrum can be
written,
\be \overline{P}_{AB}(k_i)=\frac{\Vu}{V_{s,i}}\int_{V_{s,i}}\dk\,
\left<\delta_A(\bk)\delta_{B}(-\bk)\right>  \ ,\ee
where the average is over the k-space shell $V_s$, of volume
\be V_{s,i}=\int_{k_i-\Delta k/2}^{k_i+\Delta k/2}\dk
=4\pi^2 k_i^2\Delta k\left[1+\frac{1}{12}
\left(\frac{\Delta k}{k_i}\right)^2\right]\ .\ee
The discretized form for the band-power is,
\be \overline{P}_{AB}(k)=\frac{\Vu}{N_{k}}\sum_{j=1}^{N_k}
\left<\delta_A(\bk_j)\delta_{B}(-\bk_j)\right> \ ,\ee
where $N_{k}=V_{s}(k)/V_k$ is the total number of modes in the shell.
$V_k=k_f^3$ is the fundamental $k$-space cell-volume and $k_f=2\pi/L$
is the fundamental wavemode.

Likewise, the band-power averaged covariance can be written,
\ba 
\overline{C}^d_{AB}[k_i,k_j] & \equiv &
\frac{1}{V_{s,i}V_{s,j}}
\int_{V_{s,i},V_{s,j}}\hspace{-0.3cm}\dk_1\dk_2 \, {C}^d_{AB}[\bk_1,\bk_2]
\label{eq:BinCovP}\ .\ea
To obtain the full non-Gaussian band-power covariance, one then
inserts Eq.~(\ref{eq:CovCrossPowFull}) into the above expression, and
this leads to,
\ba 
\overline{C}^d_{AB}[k_i,k_j]  & = & \frac{1}{\Vu} \overline{T}_{AABB}[k_i,k_j] \nonumber \\
& & \hspace{-2cm} +\frac{1}{N_{k}}\left[\left(\overline{P}_{AA}(k_i)+\frac{1}{\nbar_A}\right)
\left(\overline{P}_{BB}(k_j)+\frac{1}{\nbar_B}\right) \right. \nonumber \\
& & \hspace{-2cm} + \left. \left(\overline{P}_{AB}(k_i)+\left\{\frac{1}{\nbar_A}\right\}\right)
 \left(\overline{P}_{AB}(k_j)+\left\{\frac{1}{\nbar_A}\right\}\right)\right]
\delta^{K}_{k_i,k_j}\nonumber\\
& & \hspace{-2cm} + \frac{\overline{B}_{AAB}(k_i,k_j)}{N_B}+
\frac{\overline{B}_{ABB}(k_i,k_j)}{N_A} 
+\left\{\frac{2}{N_A}\overline{B}_{ABB}(k_i,k_j)\right\} \nonumber \\
& & \hspace{-2cm} +\frac{\overline{P}_{AB}[k_i,k_j]}{\nbar_A\nbar_B \Vu}
+ \left\{ 
\frac{1}{\nbar_A^2\Vu}\left[\overline{P}_{BB}[k_i,k_j]\right]
\right\} \nonumber \\
& & \hspace{-2cm} +\left\{
\frac{1}{\nbar_A\nbar_B\Vu}\left[\overline{P}_{AB}(k_i)+\overline{P}_{AB}(k_j)\right]\right\}
\nonumber \\
& & \hspace{-2cm} +\left\{\frac{1}{\nbar_A^2\Vu}\left[\overline{P}_{BB}(k_i)+\overline{P}_{BB}(k_j)\right]
\right\} + \left\{\frac{1}{\nbar_A^2\nbar_B V_{\mu}} 
\right\} \label{eq:CovCrossPowFullBin}\ .
\ea
where the bin averaged trispectrum and bispectrum are:
\ba 
\overline{T}[k_i,k_j]  & \equiv & 
\int_{V_{s,i},V_{s,j}}
\frac{\dk_1}{V_{s,i}}\frac{\dk_2}{V_{s,j}} T(\bk_1,\bk_2,-\bk_1,-\bk_2)\ ; \\
\overline{B}[k_i,k_j]  & \equiv & 
\int_{V_{s,i},V_{s,j}}
\frac{\dk_1}{V_{s,i}}\frac{\dk_2}{V_{s,j}} B(\bk_1,\bk_2,-\bk_1-\bk_2)\ ; 
\ea
and where we introduced the function,
\ba 
\overline{P}[k_i,k_j]  & \equiv & 
\int_{V_{s,i},V_{s,j}}
\frac{\dk_1}{V_{s,i}}\frac{\dk_2}{V_{s,j}} P(\left|\bk_1-\bk_2\right|)\ , \\
 & = & 
\int_{V_{s,i},V_{s,j}}
\frac{\dk_1}{V_{s,i}}\frac{\dk_2}{V_{s,j}} P(\left|\bk_1+\bk_2\right|)\ ,
\ea
We may now consider a number of interesting limiting cases of the
above expressions. Firstly, in the very large-scale limit
$\{k_i,k_j\}\rightarrow0$, and we have that
$\overline{P}\rightarrow0$, $\overline{B}\rightarrow0$ and
$\overline{T}\rightarrow0$, and the covariance becomes
\be \overline{C}^d_{AB}[k_i,k_j]   \approx  
\frac{1}{N_{k_i}}\frac{1}{\nbar_A\nbar_B} \delta^{K}_{k_i,k_j}
+\left\{\frac{1}{\nbar_A^2\nbar_B V_{\mu}}\right\}\ . 
\ee
In the small-scale limit $\{k_i,k_j\}\gg k_f=2\pi/L$, and again we have that
$\overline{P}\rightarrow0$, $\overline{B}\rightarrow0$ and
$\overline{T}\rightarrow0$, and also $N_k\gg1$, hence
\be \overline{C}^d_{AB}[k_i,k_j] \approx
\left\{\frac{1}{\nbar_A^2\nbar_B V_{\mu}}\right\} \ .\label{eq:LargekCov}\ee
The correlation matrix ${\mathcal C}$ is defined as the covariance
matrix normalized by its diagonal components, i.e.
\be {\mathcal C}^d_{AB}[k_i,k_j]=\frac{\overline{C}^d_{AB}[k_i,k_j]}{
\sqrt{\overline{C}^d_{AB}[k_i,k_i]\overline{C}^d_{AB}[k_j,k_j]}}\ ,\ee
and ${\mathcal C}[k_i,k_i]=1$ and $-1\le{\mathcal C}[k_i,k_j]\le1$.
Thus for $(i\ne j)$ and in the large-scale limit, we find
\ba 
{\mathcal C}^d_{AB}[k_i,k_j] & \approx &
\left[
\frac{(\nbar_A\Vu)^2}{N_{k_i}N_{k_j}}
+\nbar_A\Vu\left(\frac{N_{k_i}+N_{k_j}}{N_{k_i}N_{k_j}}\right)+1
\right]^{-1/2}\ .\nonumber \\
& \approx & \frac{\sqrt{N_{k_i}N_{k_j}}}{\nbar_A\Vu} \ll 1 \ ,\ea
where the second equality obtains from assuming
$\nbar_A\Vu\rightarrow\infty$. Conversely, in the small-scale limit we find
\ba {\mathcal C}^d_{AB}[k_i,k_j] & \approx & 1  .\ea
These last two expressions are important results. The first informs us
that if one computes the auto-power spectrum of a discrete sampling of
the density field, then for a standard CDM power spectrum, the
covariance matrix is diagonal on large scales provided
$\nbar\Vu\gg1$. However on small scales all of the Fourier modes
inevitably become perfectly correlated, and this is completely
independent of any Non-Gaussian terms generated through gravitational
instability. On the other hand, if there is no overlap between samples
A and B, then there will be no off-diagonal covariance, since
Eq.~(\ref{eq:LargekCov}) vanishes.

We may demonstrate these statements more clearly by taking the Gaussian 
limit of Eq.~(\ref{eq:CovCrossPowFullBin}), 
\ba 
\overline{C}^d_{AB}[k_i,k_j]  & = & 
\frac{1}{N_{k}}
\left[
\left(\overline{P}_{AA}(k_i)+\frac{1}{\nbar_A}\right)
\left(\overline{P}_{BB}(k_j)+\frac{1}{\nbar_B}\right) 
\right. \nonumber \\
& & \hspace{-2.1cm} + 
\left. 
\left(\overline{P}_{AB}(k_i)+\left\{\frac{1}{\nbar_A}\right\}\right)
 \left(\overline{P}_{AB}(k_j)+\left\{\frac{1}{\nbar_A}\right\}\right)
\right] \delta^{K}_{k_i,k_j}\nonumber\\
& & \hspace{-2.1cm} +
\frac{\overline{P}_{AB}[k_i,k_j]}{\nbar_A\nbar_B \Vu} + 
\left\{ 
\frac{1}{\nbar_A^2\Vu}\left[\overline{P}_{BB}[k_i,k_j]\right]
\right\} \nonumber \\
& & \hspace{-2.1cm} +
\left\{\frac{1}{\nbar_A^2\Vu}\left[\overline{P}_{BB}(k_i)+\overline{P}_{BB}(k_j)\right]
\right\} \nonumber \\
& & \hspace{-2.1cm} +
\left\{
\frac{1}{\nbar_A\nbar_B\Vu}\left[\overline{P}_{AB}(k_i)+\overline{P}_{AB}(k_j)\right]
\right\}+
\left\{ \frac{1}{\nbar_A^2\nbar_B V_{\mu}}\right\} 
\label{eq:CrossPowerGaussBin}\ .
\ea
For the case where samples A and B are identical, then the above
expressions reduce to,
\ba 
\overline{C}^d[k_i,k_j] \!\!\! & = & 
\frac{2}{N_{k_i}}
\left(\overline{P}(k_i)+\frac{1}{\nbar}\right)^2\delta^{K}_{k_i,k_j}\nonumber\\
& + & \hspace{0cm}
\frac{2}{\nbar^2\Vu}\left[ \overline{P}[k_i,k_j]+ \overline{P}(k_i)+ \overline{P}(k_j) \right]
+\frac{1}{\nbar^3 V_{\mu}}
\label{eq:CrossAutoPowerGaussBin}\ .
\ea
Finally, since it will be of use later, we may also take the limit 
$\nbar_A\Vu\rightarrow\infty$, giving
\ba 
\overline{C}^d_{AB}[k_i,k_j]  & = & 
\frac{1}{N_{k}}
\left[
\left(\overline{P}_{AA}(k_i)+\frac{1}{\nbar_A}\right)
\left(\overline{P}_{BB}(k_j)+\frac{1}{\nbar_B}\right) 
\right. \nonumber \\
& & \hspace{-2.1cm} + 
\left. 
\left(\overline{P}_{AB}(k_i)+\left\{\frac{1}{\nbar_A}\right\}\right)
 \left(\overline{P}_{AB}(k_j)+\left\{\frac{1}{\nbar_A}\right\}\right)
\right] \delta^{K}_{k_i,k_j} \label{eq:CrossPowerGaussBinLargeN}\ .
\ea
and for A equivalent to B,
\ba 
\overline{C}^d[k_i,k_j] \!\!\! & = & 
\frac{2}{N_{k_i}}
\left(\overline{P}(k_i)+\frac{1}{\nbar}\right)^2\delta^{K}_{k_i,k_j} \ 
\label{eq:CrossAutoPowerGaussBinLargeN}\ .
\ea

Figure~\ref{fig:CorrMatrixHH} presents the correlation matrix for the
halo-halo auto-power spectrum generated using
Eq.~(\ref{eq:CrossAutoPowerGaussBin}). In the left and right panels we
show the results for haloes with masses in the range
$M\in[1.0,2.0]\times10^{13}\Msol$ and
$M\in[1.0,2.0]\times10^{15}\Msol$, respectively. We evaluate the
average bias and halo number density within each mass bin using the
\citep{ShethTormen1999} models, and we find,
$\nbar=\left[1.87\times10^{-4},1.12\times10^{-7}\right] h^3~{\rm
  Mpc}^{-3}$ and $b=\left[1.30, 5.85\right]$, for the two bins
respectively, and we take $\Vu=(1500)^3\Mpccube$. In both cases the
matrix becomes fully correlated and the rare sample shows a much
stronger correlation on larger scales than the higher abundance lower
mass halo sample. On the other hand, if we were to compute the
correlation matrix for the cross-power spectrum of the two halo
samples, then we would predict that the correlation matrix would be
equivalent to the identity matrix.

Before we leave this section, it is interesting to note that, in the
pure Gaussian limit, i.e. $\nbar_TP_{T}\gg1$, then the fractional
variance in the cross-power is not simply dependent upon the number of
available modes, but also the cross-correlation coefficient:
$r_{AB}(k)\equiv P_{AB}(k_i)/\sqrt{P_{AA}(k_i)P_{BB}(k_i)}$. This can
be seen directly from Eq.~(\ref{eq:CrossPowerGaussBin}),
\be \left(\frac{\sigma_{\overline{P}_{AB}}}{\overline{P}_{AB}}\right)^2=\frac{1}{N_k}
\left(\frac{1}{r_{AB}^2}+1\right)\ . \label{eq:fracerr}
\ee
The corresponding expression for the auto-power spectrum is
$\overline{\sigma}_P/\overline{P}=\sqrt{2/N_k}\propto
k^{-1}\Vu^{-1/2}$. However, when $r_{AB}=1$, then there is no
difference and the fractional error scales with the survey volume in
the usual way. In \S\ref{ssec:bandpower}, we shall show that for
haloes and dark matter on the largest scales, the cross-power approach
offers only a modest improvement over the auto-power method, implying
that $r_{AB}\approx 1$.


\section{Efficiency of estimators}\label{sec:Efficiency}

\subsection{Comparing estimators}

One might ask the following question: when should one apply the
cross-power spectrum approach, instead of the usual auto-power
spectrum approach? In this section we shall attempt to answer this
question. The main advantages of the cross-power approach are most
apparent when one probes the dependence of a given sample of tracers
as a function of some physical parameter, i.e. the luminosity
dependence of clustering or the mass dependence of the clustering of
clusters. This statement can be more directly quantified if we
consider the concept of estimator efficiency.

If we have two unbiased estimators ${\mathcal E}_1$ and ${\mathcal
  E}_2$, then the most efficient estimator of the two, is said to be
the one with the smallest variance: i.e. if ${\rm
  Var}[\mathcal{E}_1]<{\rm Var}[\mathcal{E}_2]$, then $\mathcal{E}_1$
will be considered to be a more efficient estimator than
$\mathcal{E}_2$ \citep{Barlow1989}. We need to modify this concept slightly since in
comparing the cross- and auto-power spectra we are not estimating the
same thing, owing to the clustering bias. Instead we shall define the
effective efficiency of the estimator through the signal-to-noise
(hereafter $\SN$) ratio: i.e. ${\mathcal E}_1$ will be considered to
be a more efficient estimator than ${\mathcal E}_2$ if ${\mathcal
  E}_1/\sqrt{\rm Var[{\mathcal E}_1]}>{\mathcal E}_2/\sqrt{\rm
  Var[{\mathcal E}_2]}>$. Or in other words the estimator with the
largest $\SN$ will be the most efficient estimator.

On taking the limit $\nbar_A\Vu\rightarrow \infty$ for
Eqns~(\ref{eq:CrossPowerGaussBin}) and
(\ref{eq:CrossAutoPowerGaussBin}), the covariance matrices are
diagonal and so we may write the $\SN$ for the auto- and cross-power
spectra as:
\ba 
\frac{\left(\SN\right)_{jj}^2}{N_k} \hspace{-0.2cm} &  = \hspace{-0.2cm} & 
\frac{1}{2}\frac{\gamma_{j}^2}{\left[1+\gamma_{j}^2\right]}
\label{eq:SNBB}\ ; \\
\frac{\left(\SN\right)_{ij}^2}{N_k} \hspace{-0.2cm} & = \hspace{-0.2cm} & 
\left[\frac{\gamma_{i}\gamma_{j}r_{ij}^2}
{
 \left(\gamma_{i}+1\right)\left(\gamma_{j}+1\right)+(\sqrt{\gamma_{i}\gamma_{j}}r_{ij}
+\left\{\delta\right\})^2
}
\right]
\label{eq:SNAB}\ 
\ea
where we have introduced the following quantities,
\ba 
\gamma_{i} & \equiv & \nbar_i P_{ii} \ ; \\
r^2_{ij} & = & P^2_{ij}/P_{ii}P_{jj} \ ; \\
\delta^2 & \equiv & \nbar_j/\nbar_i\ .
\ea
where we have taken the index $i$ to denote the high density sample
$A$, and $j$ to denote the low density sample B.
Taking the ratio of the above expressions, gives us a simple test for
the relative efficiency of the estimators,
\ba
\frac{
\left(\SN\right)_{ij}^2}{\left(\SN\right)_{jj}^2} & = &
2r_{ij}^2\frac{\gamma_i}{\gamma_j} \nonumber \\
& \times & 
\left[\frac{(1+\gamma_j)^2}{(1+\gamma_i)(1+\gamma_j)+(\sqrt{\gamma_{i}\gamma_{j}}r_{ij}
+\left\{\delta\right\})^2}\right]
\label{eq:SN2}\ \ea
and we see that the relative efficiency does not depend explicitly on
the number of available modes, nor the survey volume.

To proceed further we must specify samples $i$ and $j$ in more
detail. Let us consider the case where sample $i$ is obtained from a
set of unbiased high density objects and where sample $j$ is obtained
from a set of highly biased but rare objects.  For this situation we
have, $\nbar_i\gg\nbar_j$. Hence, $\delta\rightarrow 0$.  Further, we
shall assume that $\gamma_i\gg\gamma_j$. Hence, Eq.~(\ref{eq:SN2})
simplifies to,
\be\frac{ \left(\SN\right)_{ij}^2} {\left(\SN\right)_{jj}^2} 
\approx
2r_{ij}^2 \left[\frac{\gamma_j^2+2\gamma_j+1}{\gamma_j^2(1+r_{AB}^2)+\gamma_j}\right]\
.\ee
On assuming that $r_{AB}\approx1$, we finally find that
\be
\frac{
\left(\SN\right)_{ij}^2}
{\left(\SN\right)_{jj}^2} \approx 
\left[\frac{2\gamma_j^2+4\gamma_j+2}{2\gamma_j^2+\gamma_j}\right]
> 1\ .\ee
This means that for examining the clustering properties of rare
samples of objects, it is more efficient to cross-correlate them with
a high-density sample, rather than to compute their auto-power
spectrum.


\subsection{Example: Improving estimates of cluster bias}

\begin{figure}
\centerline{
\includegraphics[width=7.5cm]{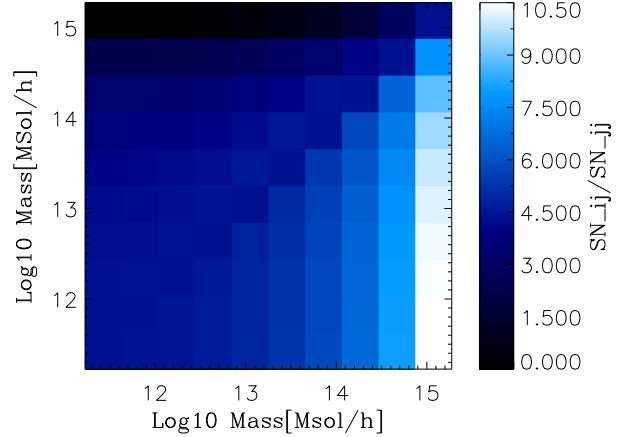}}
\caption{\small{Relative Signal-to-Noise ratio matrix
    $(\SN)_{ij}/(\SN)_{jj}$ of the cross power spectra of cluster
    samples in mass bin $i$ (y-axis) and mass bin $j$
    (x-axis).\label{fig:SNRatio}}}
\end{figure}


Let us now provide a more concrete example. Consider a sample of dark
matter clusters and suppose that we have both the redshift and an
unbiased estimate of the cluster mass, i.e. through either weak
lensing, the Sunyaev-Zel'Dovich effect etc., and that the clusters
span the mass range $M\in\left[10^{11},5\times10^{15}\right]\Msol$. We
are interested in exploring the bias as a function of cluster mass,
perhaps for use in constraining primordial Non-Gaussianity as in
\citet{Slosaretal2008}. The sample may be sub-divided into mass bins
and one may measure the auto-power spectrum of each mass bin and also
the cross-power spectra of the different mass bins.

Figure~\ref{fig:SNRatio} shows how the relative $\SN$ as given by
Eq.~(\ref{eq:SN2}), varies as a function of the mass bins $i$, and
$j$. Note that in the figure $i$ and $j$ represent the rows and
columns of the matrix, respectively. When $i<j$, then we find that
there are significant advantages to be gained from computing the
cross-power spectrum as opposed to the auto-power spectrum especially
for the case of high mass haloes. For the case where $i>j$, then,
naturally, the cross-power spectra are not optimal measures compared
to the auto-spectrum.


\section{Covariance of the cross-correlation function}\label{sec:CovCorr}

As a corollary to our study of the cross-power spectrum, we extend our
analysis to encompass the covariance of the cross-correlation
function. We note that the auto-correlation covariance of dark matter
and haloes on scales relevant for the Baryonic Acoustic Oscillations
($r\sim100\Mpc$), was recently investigated in detail by
\citet{Smithetal2008a} and \citet{Sanchezetal2008}. Here we perform a
similar study for the cross-correlation function.  

In direct analogy with the analysis of power spectrum band-powers, we
may define the band averaged cross-correlation function as,
\be \overline{\xi}^{AB}(\br_i) = \frac{1}{V_s(r_i)}\int_{V_s(r_i)} \dr\,
\xi^{AB}(r) = \int \frac{\dk}{(2\pi)^3} P^{AB}(k) \overline{j_0}(kr_i)
\label{eq:bandxi}\ee
where $V_s$ is the radial shell of thickness $\Delta r$, over which
the average is performed and this has volume,
\be V_{s,i}=4\pi r_i^2\Delta r \left[1+\frac{1}{12}
\left(\frac{\Delta r}{r_i}\right)^2\right]\ .\ee
For the second equality in Eq.~(\ref{eq:bandxi}), we have made use of
the fact that $\xi\Leftrightarrow P$ are Fourier dual, and we have
defined the zeroth order bin-averaged spherical Bessel function as,
\be \overline{{j}_0}(kr_i)\equiv 
\frac{\left.r^2j_1(kr)\right|_{r_1}^{r_2}}
{r_i^2k\Delta r\left[1+\frac{1}{12}\left(\frac{\Delta r}{r_i}\right)^2\right]} \ \ \ ;
\left\{\begin{array}{l}
r_2 = r_i+\Delta r/2 \\
r_1 = r_i-\Delta r/2\end{array}\right.
\ee
with $j_1(x) \equiv \sin x/x^2-\cos x/x$ being the 1st order spherical
Bessel function. Similar to the bin averaged covariance for the power
(c.f. Eq.~\ref{eq:BinCovP}), we may also define the bin averaged
cross-correlation covariance between bins $i$ and $j$,
\ba 
\overline{C}^d_{\xi^{AB}} & \equiv & 
{\rm Cov}\left[\overline{\xi}^{AB}_i,\overline{\xi}^{AB}_j\right] \nonumber\\
& = & \frac{1}{V_{s,i}V_{s,j}}
\int_{V_{s,i},V_{s,j}}\dr_1\dr_2 \, C^d_{\xi^{AB}}
\ .\label{eq:BinCovXi}\ea
where $C^d_{\xi^{AB}}={\rm
Cov}\left[\xi^{AB}(\br_1),\xi^{AB}(\br_2)\right]$. On inserting our
expression for the bin averaged correlation function then we may
rewrite the above expression as,
\be 
\overline{C}^d_{\xi^{AB}} =  \int 
\frac{\dk_1}{(2\pi)^3} 
\frac{\dk_2}{(2\pi)^3} 
\overline{j_0}(k_1r_i) \overline{j_0}(k_2r_j) C^d_{P_{AB}}\ .
\label{eq:BinCovXi2}\ee
Thus, the cross-power covariance also gives us the cross-correlation
covariance. It should also be noted that, even if $C^d_{P_{AB}}$ is
diagonal, then $\overline{C}^d_{\xi^{AB}}$ is not, since the spherical
Bessel functions in the integrand effectively smooth the information
across different scales. 

The full non-Gaussian contributions to the correlation covariance can
be calculated by substituting Eq.~(\ref{eq:CovCrossPowFull}) into the
above expression. On taking the continuum limit for the Kronecker
delta symbols, i.e.  $\delta^{K}_{\bk_1,\bk_2}\rightarrow
\delta^{D}(\bk_1-\bk_2)(2\pi)^3/\Vu$, rewriting the spherical Bessel
functions as,
\be j_0(kr)=\frac{1}{4\pi}\int d\Omega_{\br} \exp(-i\bk\cdot\br)\ ,\ee
and using the Fourier relations between the $N$-point correlation
functions and poly-spectra, we then find that
\ba 
\overline{C}^d_{\xi^{AB}} \hspace{-0.2cm} 
& = & \hspace{-0.2cm}  
\int_{V_{s,i}V_{s,j}} \frac{\dr_1}{V_{s,i}}\frac{\dr_2}{V_{s,j}} 
\int \frac{\dy}{\Vu} 
\eta_{AABB}(\br_1+\by,\br_2,\by) \nonumber \\
&  + & \hspace{-0.2cm}  
\int_{V_{s,i}V_{s,j}} \frac{\dr_1}{V_{s,i}}
\frac{\dr_j}{V_{s,j}} \int \frac{\dy}{\Vu} 
\left[\xi_{AA}(\by)\xi_{BB}(\br_1+\br_2+\by)\right. \nonumber \\
& + & \left. \frac{}{}\xi_{AB}(\by)\xi_{AB}(\br_1+\br_2+\by)\right]\nonumber \\
& + & \hspace{-0.2cm}  
\frac{1}{\Vu}\int_{V_{s,i}V_{s,j}} \frac{\dr_1}{V_{s,i}}
\frac{\dr_j}{V_{s,j}} 
\left[\frac{1}{\nbar_B}\xi_{AA}(\br_1+\br_2)\right.\nonumber \\
& + & \hspace{-0.2cm}  
\left. \frac{1}{\nbar_A}\xi_{BB}(\br_1+\br_2)
+\left\{\frac{2}{\nbar_A}\xi_{AB}(\br_1+\br_2)\right\}
\right] \nonumber \\ 
& + & \hspace{-0.2cm}  
\left[\frac{1}{\nbar_{A}\nbar_B}+\left\{\frac{1}{\nbar_A^2}\right\} \right]
\frac{\delta^K_{i,j}}{\Vu V_{s}(i)}+ \frac{\overline{\zeta}_{AAB}(r_i,r_j)}{N_B} \nonumber \\
& + & \hspace{-0.2cm}  
\frac{1}{N_A}\left[\overline{\zeta}_{ABB}(r_i,r_j)+
\left\{ 2\overline{\zeta}_{ABB}(r_i,r_j)\right\} \right]\nonumber \\
& + & \hspace{-0.2cm}  
\frac{\overline{\xi}_{AB}(r_j)}{\nbar_A\nbar_B\Vu V_{s,i}}\delta^{K}_{i,j} 
+ \left\{ 
\frac{\overline{\xi}_{BB}(r_i)}{\nbar_A^2\Vu V_{s,i}}\delta^{K}_{i,j} 
\right\}\nonumber \\
& + & \hspace{-0.2cm}  
\left\{ 
\frac{1}{\nbar_A\nbar_B\Vu} 
\left[\frac{\overline{\xi}_{AB}(r_i)}{V_{s,j}}\delta^{K}_{j,1} +
\frac{\overline{\xi}_{AB}(r_j)}{V_{s,i}}\delta^{K}_{i,1} \right]
\right\}\nonumber \\
& + & \hspace{-0.2cm}  
\left\{ \frac{1}{\nbar_A^2\nbar_B\Vu }
\frac{\delta^{K}_{i,1}}{V_{s,i}}
\frac{\delta^{K}_{j,1}}{V_{s,j}}
\right\}
\ea
Again we may take the Gaussian ($\eta=\zeta=0$) limit of the full
expression and we find
\ba \overline{C}^d_{\xi^{AB}} & = & \frac{1}{\Vu}\int
\frac{\dk}{(2\pi)^3} \overline{j_0}(kr_i) \overline{j_0}(kr_j)
\nonumber \\
&  \times & \left[
\left(P_{AA}(k_1)+\frac{1}{\nbar_A}\right)
\left(P_{BB}(k_1)+\frac{1}{\nbar_B}\right) \right. \nonumber \\ 
& + & \left. \left(P_{AB}(k_1) +\left\{\frac{1}{\nbar_A}\right\}\right)^2\right]
\nonumber \\ 
\hspace{-0.2cm}  
& + & \frac{\overline{\xi}_{AB}(r_j)}{\nbar_A\nbar_B\Vu V_{s,i}}\delta^{K}_{i,j} 
+ \left\{ 
\frac{\overline{\xi}_{BB}(r_i)}{\nbar_A^2\Vu V_{s,i}}\delta^{K}_{i,j} 
\right\}\nonumber \\
& + & \left\{ 
\frac{1}{\nbar_A\nbar_B\Vu} 
\left[
\frac{\overline{\xi}_{AB}(r_i)}{V_{s,j}}\delta^{K}_{j,1} +
\frac{\overline{\xi}_{AB}(r_j)}{V_{s,i}}\delta^{K}_{i,1} \right]
\right\}\nonumber \\
& + & \left\{ \frac{1}{\nbar_A^2\nbar_B\Vu }
\frac{\delta^{K}_{i,1}}{V_{s,i}}
\frac{\delta^{K}_{j,1}}{V_{s,j}}
\right\} \ .
\ea
The first term is the usual Gaussian plus Poisson expression and this
leads to off-diagonal covariance through the spherical Bessel
functions. The second and third terms contribute only to the diagonal
variance, however the fourth and fifth terms contribute to the
off-diagonal variance along lines of zero lag and the last contributes
only to the zero lag term. Therefore in the Gaussian limit, whilst the
covariance matrix for the correlation functions is in general
non-diagonal, the terms associated with the Poisson noise that lead to
off-diagonal terms in the power spectrum covariance, do not generate
off-diagonal covariance in the correlation function. However, in the
more general case we see that additional off-diagonal terms can be
generated when we have non-zero $\eta$ and $\zeta$. Furthermore, for
the case of the cross-correlation function of a non-overlapping
samples, then all of the terms in curly brackets vanish, and the
covariance is significantly reduced.

Finally, on taking the limit $\nbar_A\Vu\rightarrow\infty$ the
covariance between band averages of the cross-correlation function
reduces to,
\ba \overline{C}^d_{\xi^{AB}} 
& \approx & \frac{1}{\Vu}\int \frac{\dk}{(2\pi)^3} 
\overline{j_0}(kr_i) \overline{j_0}(kr_j) \Gamma(k)\nonumber \\
& + & \frac{\delta^{K}_{i,j}}{\Vu V_{s}(r_i)}
\left[\frac{1}{\nbar_A\nbar_B}+\left\{\frac{1}{\nbar_A^2}\right\}\right]\ ,
\ea
and where we introduced the useful function
\[ 
\Gamma(k) = P_{AA}(k_1)P_{BB}(k_1)+
\left[\frac{P_{AA}(k_1)}{\nbar_B}+\frac{P_{BB}(k_1)}{\nbar_A}\right]
\]
\be
\hspace{1cm}  
+P_{AB}^2(k_1)+\left\{\frac{2P_{AB}(k_1)}{\nbar_A}\right\}\ .\ee
Lastly, in the limit where $\nbar_A\equiv\nbar_B$, we recover the
usual expression in the Gaussian limit
\citep{Smithetal2008a,Sanchezetal2008}.


\begin{table*}
\centering{
\caption{Parameters for the {\tt zHORIZON} simulations  -- Columns are:
density parameters for matter, dark energy and baryons; the equation
of state parameter for the dark energy $P_{w}=w\rho_{\rm w}$;
normalization and primordial spectral index of the power spectrum;
dimensionless Hubble parameter $H_{0}=h 100 [\Hub]$; number of
particles, box size, particle mass, number of realizations, and total simulation
volume, respectively.}
\label{tab:params}
\begin{tabular}{c|cccccccccccc}
\hline 
Simulation & $\Omega_m$ & $\Omega_{w}$ & $\Omega_b$ & $w_0$  &  
$\sigma_8$  & $n$ &  $h$ & $N$ & $L[\Mpc]$ & $m_{\rm p}[\Msol] $ & $N_{\rm sim}$ & $V_{\rm tot}[\Gpccube]$\\
\hline
{\tt zHORIZON}   & 0.25\ &  0.75 & 0.04 &  -1  &  0.8  & 1.0 & 0.7
& $750^3$ & 1500.0 & $5.55\times10^{11}$ & 30 & 101.25
\end{tabular}}
\end{table*}


\section{Comparison with $N$-body simulations}\label{sec:simulations}

In this section we compare the counts-in-cells predictions for the
covariance matrices of the mass-mass, halo-mass and halo-halo power
spectra and correlation functions with results from a large ensemble
of numerical simulations.


\subsection{The {\tt zHORIZON} simulations}

The Z\"urich Horizon, ``{\tt zHORIZON}'', simulations are a large
ensemble of pure cold dark matter $N$-body simulations ($N_{\rm
  sim}=30,\, N_{\rm part}=750^3)$, performed at the University of
Z\"urich on the {\tt zBOX2} and {\tt zBOX3} super-computers. The
specific aim for these simulations is to provide high precision
measurements of cosmic structures on the scales of the order
$\sim100\Mpc$ and to also provide insight into the rarest fluctuations
within the LCDM model that we should expect to find within the
observable universe -- the Horizon Volume. 

Each numerical simulation was performed using the publicly available
{\tt Gadget-2} code \citep{Springel2005}, and followed the nonlinear
evolution under gravity of $N$ equal mass particles in a comoving cube
of length $L$. All of the simulations were run within the same
cosmological model, and the particular choice for the parameters was
inspired by results from the WMAP experiment
\citep{Spergeletal2003,Spergeletal2007,Komatsuetal2008} -- the
parameters for the simulations are listed in
Table~\ref{tab:params}. The transfer function for the simulations was
generated using the publicly available {\tt cmbfast} code
\citep{SeljakZaldarriaga1996,Seljaketal2003b}, with high sampling of
the spatial frequencies on large scales. Initial conditions were lain
down at redshift $z=50$ using the serial version of the publicly
available {\tt 2LPT} code \citep{Scoccimarro1998,Crocceetal2006}.

Dark matter halo catalogues were generated for all snapshots of each
simulation using the Friends-of-Friends (FoF) algorithm
\citep{Davisetal1985}, with the linking-length parameter set to the
standard $b=0.2$ -- $b$ is the fraction of the inter-particle
spacing. For this we used the fast parallel {\tt B-FoF} code, kindly
provided by V.~Springel. The minimum number of particles for which an
object was considered to be a bound halo, was set to 30
particles. This gave a minimum host halo mass of $\sim10^{13}
M_{\odot}/h$.


\begin{figure*}
\centerline{
\includegraphics[width=13cm]{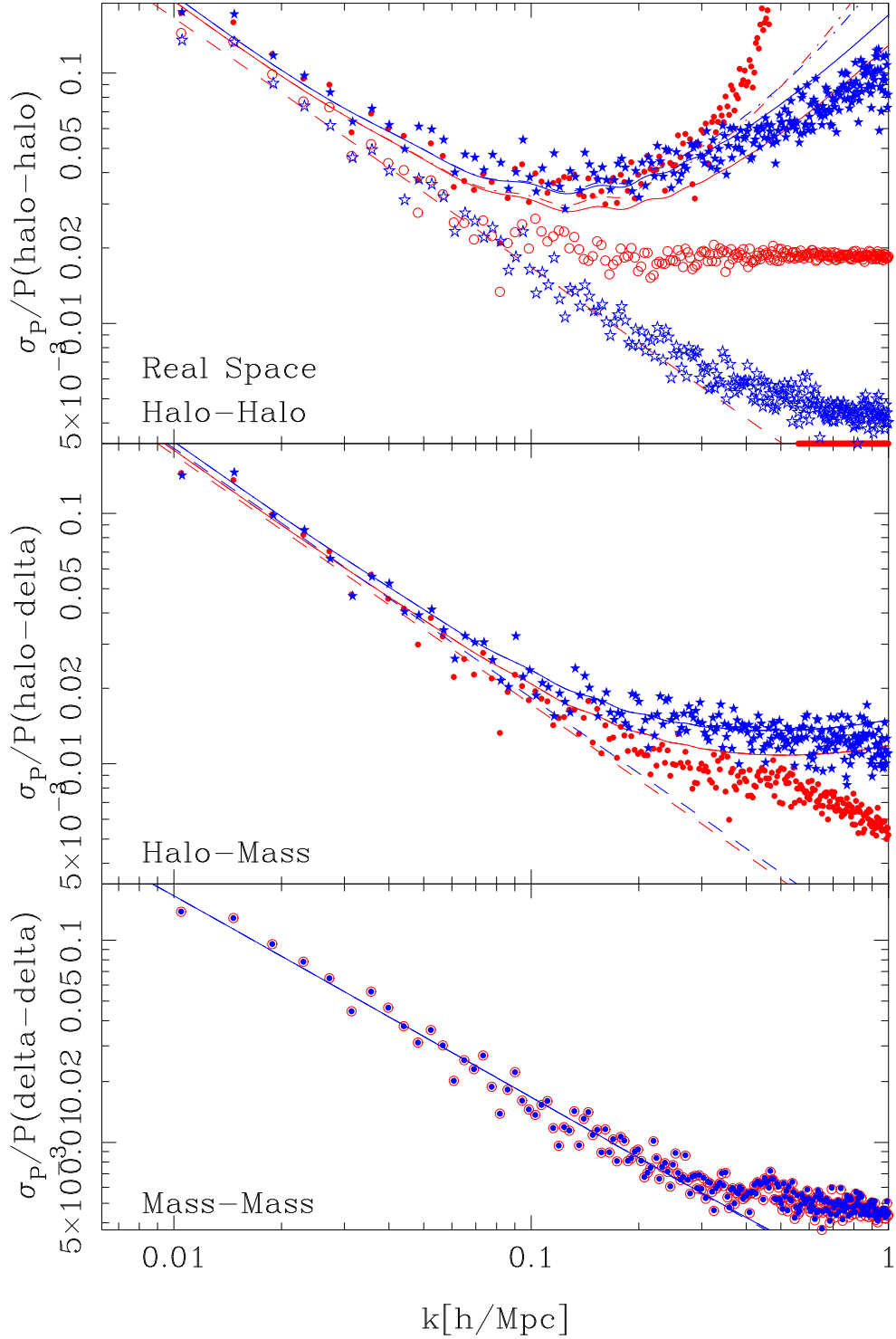}}
\caption{\small{Comparison of the fractional variance in the halo and
    mass power spectra measured from the {\tt zHORIZON} simulations
    with theoretical predictions.  The three panels show the square
    root of the bin averaged diagonal elements of the covariance
    matrix, ratioed to the mean power in the bin as a function of the
    spatial frequency. From top to bottom, the panels show results for
    the halo-halo, halo-mass and mass-mass power spectra. In all
    panels, solid points denote results obtained after a standard shot
    noise subtraction, and corresponding open points denote results
    prior to shot-noise subtraction. Dash lines represent the pure
    Gaussian predictions. Solid lines denote theoretical predictions
    from the Gaussian plus standard Poisson noise theory.  Dot-Dash
    lines denote the results from Eqs~(\ref{eq:CrossPowerGaussBin})
    and (\ref{eq:CrossAutoPowerGaussBin}). In the top two panels, the
    (red) point symbols and (blue) star symbols denote haloes with
    masses in the range $(M>1.0\times10^{14}\Msol)$ and
    $(1.0\times10^{13}<M<2\times10^{13}\Msol)$, respectively.}
\label{fig:PowError}}
\end{figure*}


\subsection{Results: band-power variances}\label{ssec:bandpower}

Figure~\ref{fig:PowError} shows the results for the mean fractional
error in the mass-mass (bottom panel), halo-mass (middle panel), and
halo-halo (top panel) power spectra, as measured from the {\tt
  zHORIZON} simulations. The spectra were estimated for each
simulation using the standard methods
\citep{Smithetal2003,Jing2005,Smithetal2008b}: particles and halo
centres were interpolated onto a $1024^3$ cubical mesh, using the CIC
algorithm \citep{HockneyEastwood1988}; the Fast Fourier Transform of
the discrete mesh was computed using the FFTW libraries \citep{FFTW};
the power in each Fourier mode was estimated and then corrected for
the CIC charge assignment; these estimates were then bin averaged in
spherical shells of thickness the fundamental frequency.

The halo-halo and halo-mass spectra were estimated for six bins in
halo mass. The thickness of the mass bins was determined by estimating
the ${\SN}$ in each bin, and demanding that it should be in excess of
20. In the figure we show the errors for an experiment of volume
$\sim3.4\Gpccube$. For clarity, we only present results for the
highest mass bin ($M> 10^{14}\Msol$, red point symbols) and for the
lowest mass bin ($10^{13}\Msol<M<1.38\times 10^{13}\Msol$, blue star
symbols) in our sample. The mean number densities in these bins are
$\nbar_{\rm h}=\{2.42,8.01\}\times10^{-5}\Mpccube$, respectively. The
mass-mass and halo-halo power spectra were both corrected for
shot-noise by subtraction of $1/\nbar=\Vu/N$ and $1/\nbar_{\rm
h}=\Vu/N_{\rm halo}$, respectively.

In the figure, the results for the shot-noise corrected and
uncorrected spectra are represented as filled and empty symbols,
respectively.  The halo bias parameters were estimated from the
cross-power and the shot-noise corrected auto-power spectrum ${\bf
  b}=(b^{\rm h\delta}_{\rm NL},b^{\rm hh}_{\rm NL})$ following the
method in \citet{Smithetal2007}. The measured values were found to be
${\bf b}=(2.803\pm0.015,3.110\pm0.015)$ and ${\bf
  b}=(1.208\pm0.010,1.479\pm0.011)$ for the highest and lowest mass
bins, respectively.  These estimates of the bias were used along with
the ensemble average number densities in the mass bin to generate the
theoretical predictions for the signal and its variance.

Considering Fig.~\ref{fig:PowError} in more detail, we note that on
the largest scales, $k\approx0.01\kMpc$, the amplitudes of the
fractional variances for all spectra are roughly equivalent. For the
auto-spectra this agreement is simply a consequence of the fact that
when the signal is dominated by the sample variance, the fractional
errors in the spectra scale as $\sigma_{P}/P = (2/N_{k})^{1/2}\propto
\Vu^{-1/2}$ (dashed lines in the figure). However as we noted earlier,
for the cross-spectrum, this near agreement also implies $r_{AB}\approx
1$. 

For the matter power spectrum (bottom panel in
Fig.~\ref{fig:PowError}), we see that this simple scaling appears to
be preserved all the way to $k\approx0.2\kMpc$, and here the errors
are of the order $1\%$ for this volume.  The scaling at this point is
broken and there is an excess of variance. This excess is not
explained by the simple addition of the usual Poisson sampling error
term (c.f. Eq.~\ref{eq:CrossAutoPowerGaussBinLargeN}), nor by the
addition of the extra shot-noise terms from the full counts-in-cells
covariance (c.f. Eq.~\ref{eq:CrossAutoPowerGaussBin}). However, in
making these predictions, we have ignored all sources of variance
generated through the nonlinear gravitational mode-coupling and it is
likely that the excess error can be attributed to these
\citep{Scoccimarroetal1999b,MeiksinWhite1999,ScoccimarroSheth2002,RimesHamilton2006,Hamiltonetal2006,Anguloetal2008a,Takahashietal2009}.


\begin{figure*}
\centerline{ 
\includegraphics[width=8.5cm]{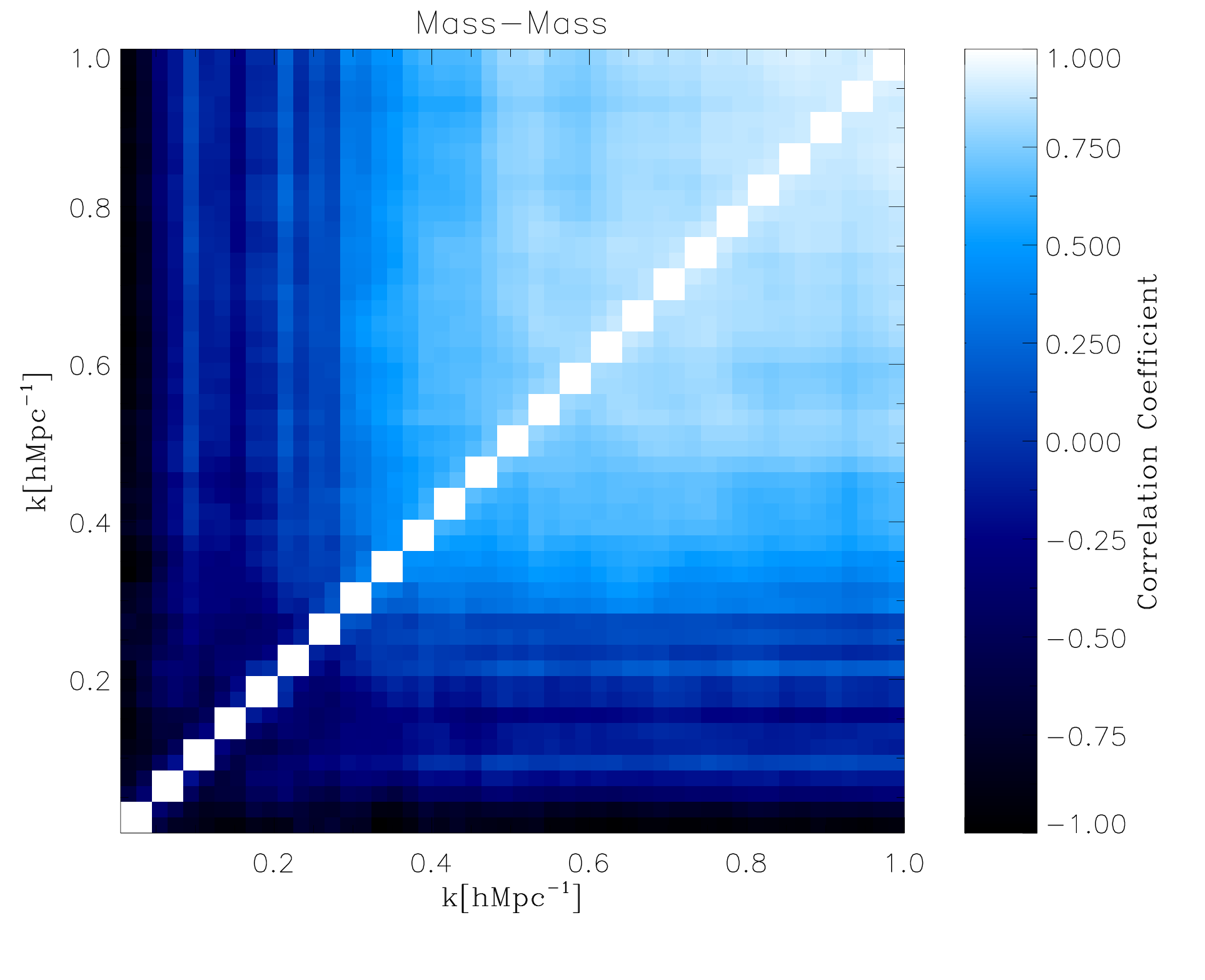}
\includegraphics[width=8.5cm]{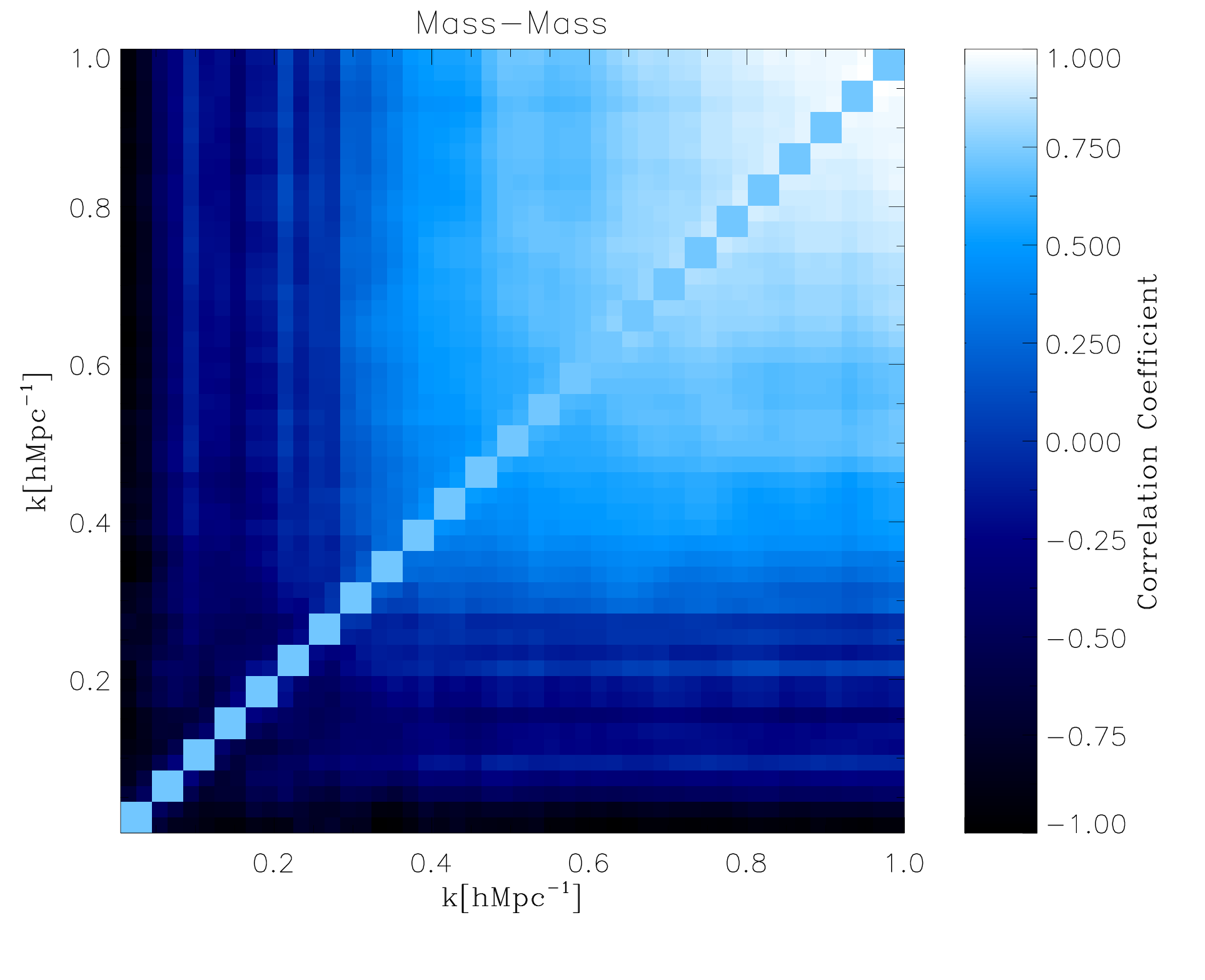}}
\caption{\small{Mass-Mass power correlation matrices measured from the
    {\tt zHORIZON} simulations. {\em Left panel:} result before
    shot-noise correction. {\em Right panel:} result after shot-noise
    correction.}
\label{fig:CovarianceMM}}
\end{figure*}


Considering the halo-mass cross-power spectra (middle panel in
Fig.~\ref{fig:PowError}), we find that the scaling with the number of
modes is broken on slightly larger scales than for matter
$(k\sim0.1\kMpc)$.  At this point the fractional error is of the order
$\sim2\%$.  However, this time the increase in the error appears to be
qualitatively described by Eq.~(\ref{eq:CrossPowerGaussBinLargeN},
solid line), although the error in the high-mass sample (red empty and
filled circles) is slightly overpredicted. The additional source of
variance in Eq.~(\ref{eq:CrossPowerGaussBin}) does not change the
predictions in any noticeable way. On smaller scales, $(k>0.1\kMpc)$,
the fractional error drops to $\sim1\%$, and is only slightly larger
than the error in the mass-mass spectrum. The excess theoretical error
suggests that haloes and dark matter are not independent samples (as
in sampling case i from \S\ref{ssec:twotracers}), more that haloes are
some `special' sub-sampling of the mass (similar to case ii), since we
expect the Gaussian error to be an underestimate. This leads us to
speculate that the halo-mass spectra also require a shot-noise
correction.

Considering the halo-halo spectra (top panel of
Fig.~\ref{fig:PowError}), we show results obtained with (solid
symbols) and without (open symbols) the standard shot-noise
subtraction. This clearly demonstrates the importance of this
correction for this sample.  In the case of the uncorrected spectra,
it appears that the errors follow the scaling with the number of modes
to high wavenumbers $(k\sim0.1\kMpc)$, where the error is of the
order $\sim2\%$.  In addition we see that the standard theoretical
predictions from Eq.~(\ref{eq:CrossAutoPowerGaussBinLargeN})
significantly over-predict the error, especially for the low-mass halo
sample. However, after shot-noise subtraction, the sample variance
scaling is actually broken on larger scales than for the
cross-spectra, and the fractional error is of the order
$\sim4-5\%$. Somewhat surprisingly, these simple theoretical
predictions provide a reasonable description of the variance and, as
for the case of the matter-matter power spectrum, are an
underestimate. If we now include the additional sources of variance
from the full counts-in-cells covariance, as given by
Eq.~(\ref{eq:CrossAutoPowerGaussBin}), then we now see that there is a
significant increase in the errors for scales $k>0.1\kMpc$. We have
again neglected the gravitational model coupling variances, but it
appears that most of the shape of this distribution is well captured
by the non-Gaussianity of the sampling procedure. On comparison with
\citet{Anguloetal2008a}, we find a slight disagreement, in that the
Gaussian plus Poisson sampling model appears in reasonable agreement
with the measurements.

Finally, we emphasize the fact that the fractional errors associated
with the cross-power spectra are more than a factor $\sim2$ times
smaller than the corresponding errors for the halo auto-spectra on
scales $k\sim0.1\kMpc$. Thus for experiments that wish to measure, for
instance, galaxy bias as a function of luminosity, halo mass or galaxy
type, then one may gain a significant increase in $\SN$ through use of
the cross-correlation approach. The caveat being that the off-diagonal
errors of the covariance matrix of the cross-power spectrum should be
small. We shall now explore this issue.


\begin{figure*}
\centerline{ 
\includegraphics[width=8.5cm]{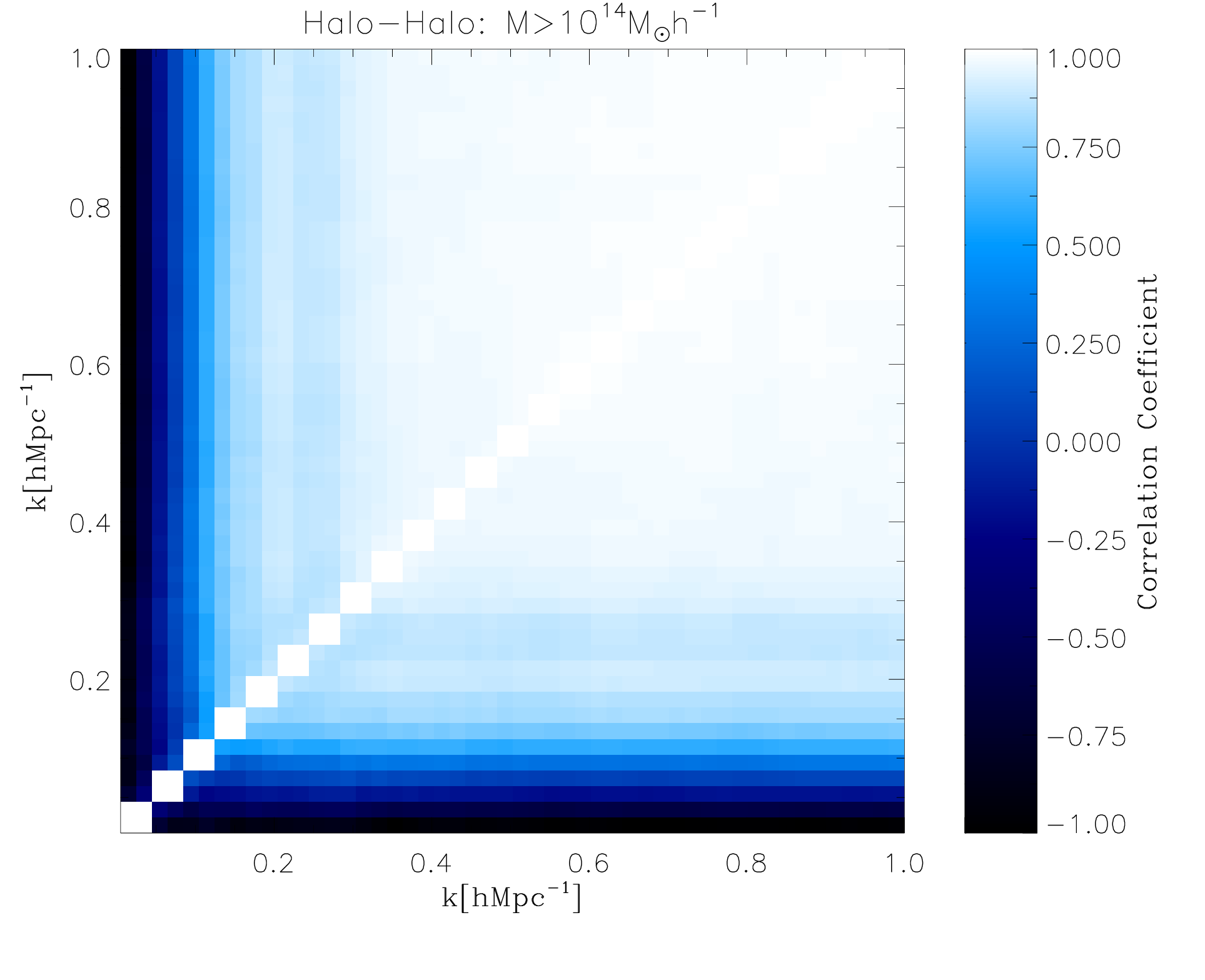}
\includegraphics[width=8.5cm]{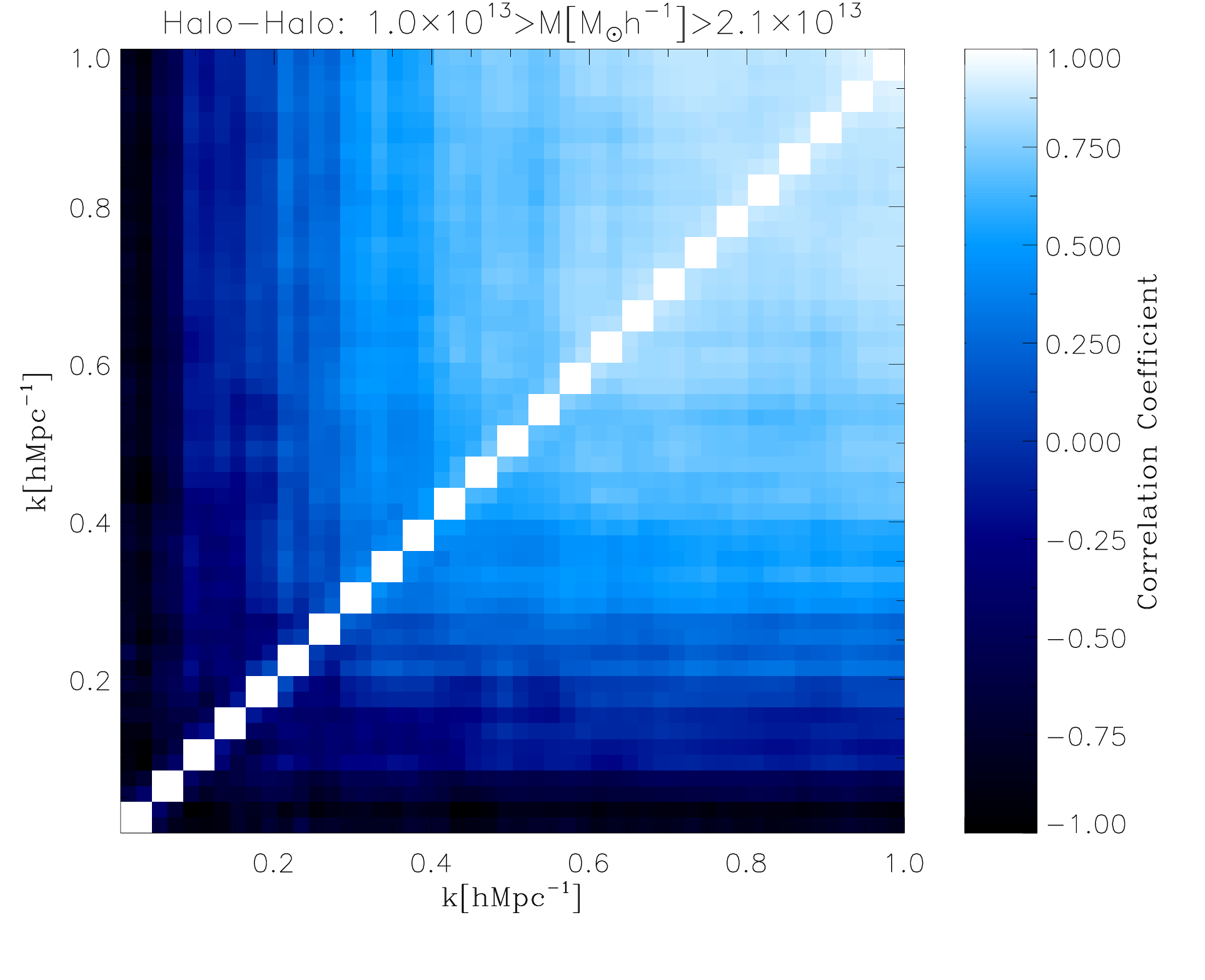}}
\centerline{ 
\includegraphics[width=8.5cm]{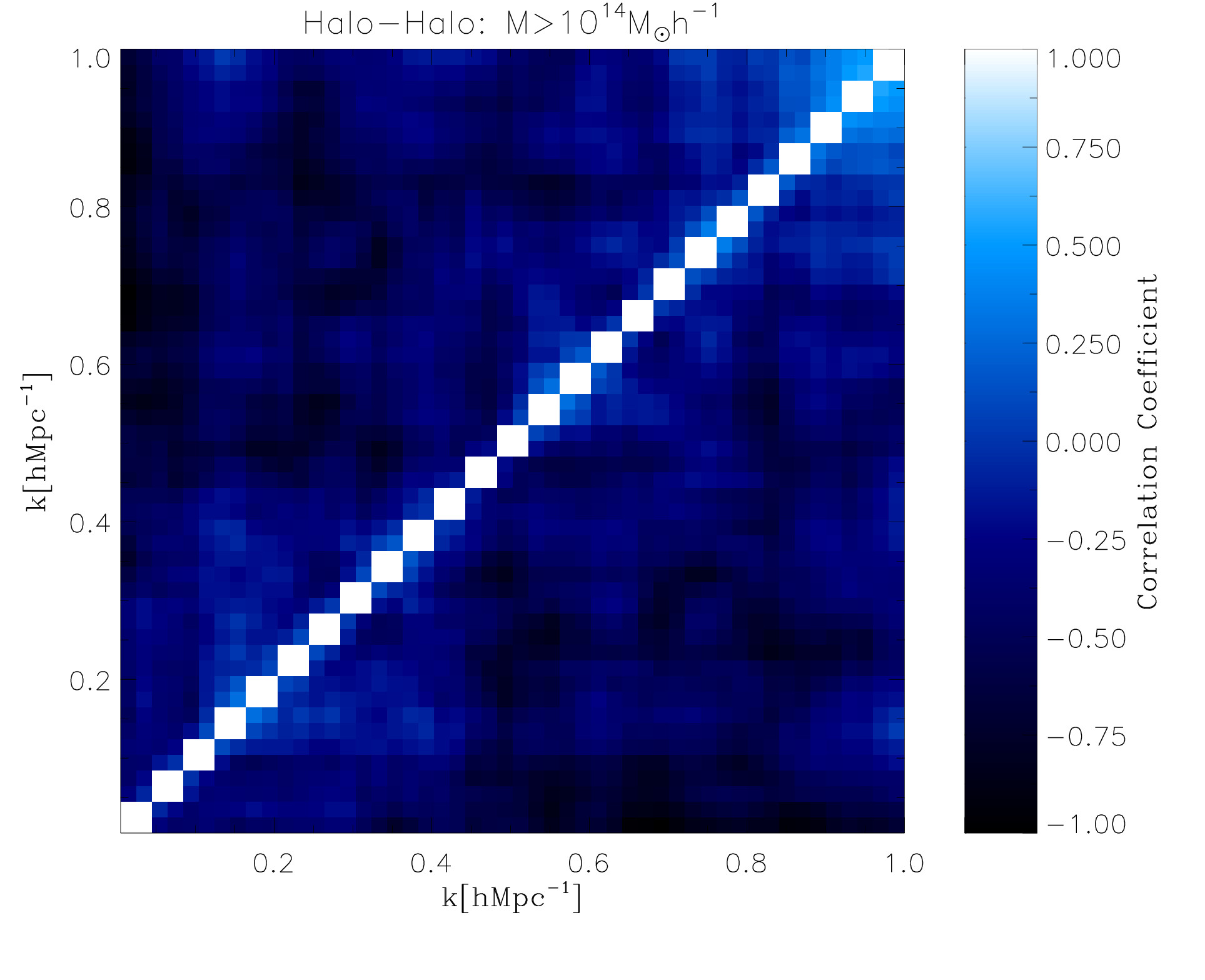}
\includegraphics[width=8.5cm]{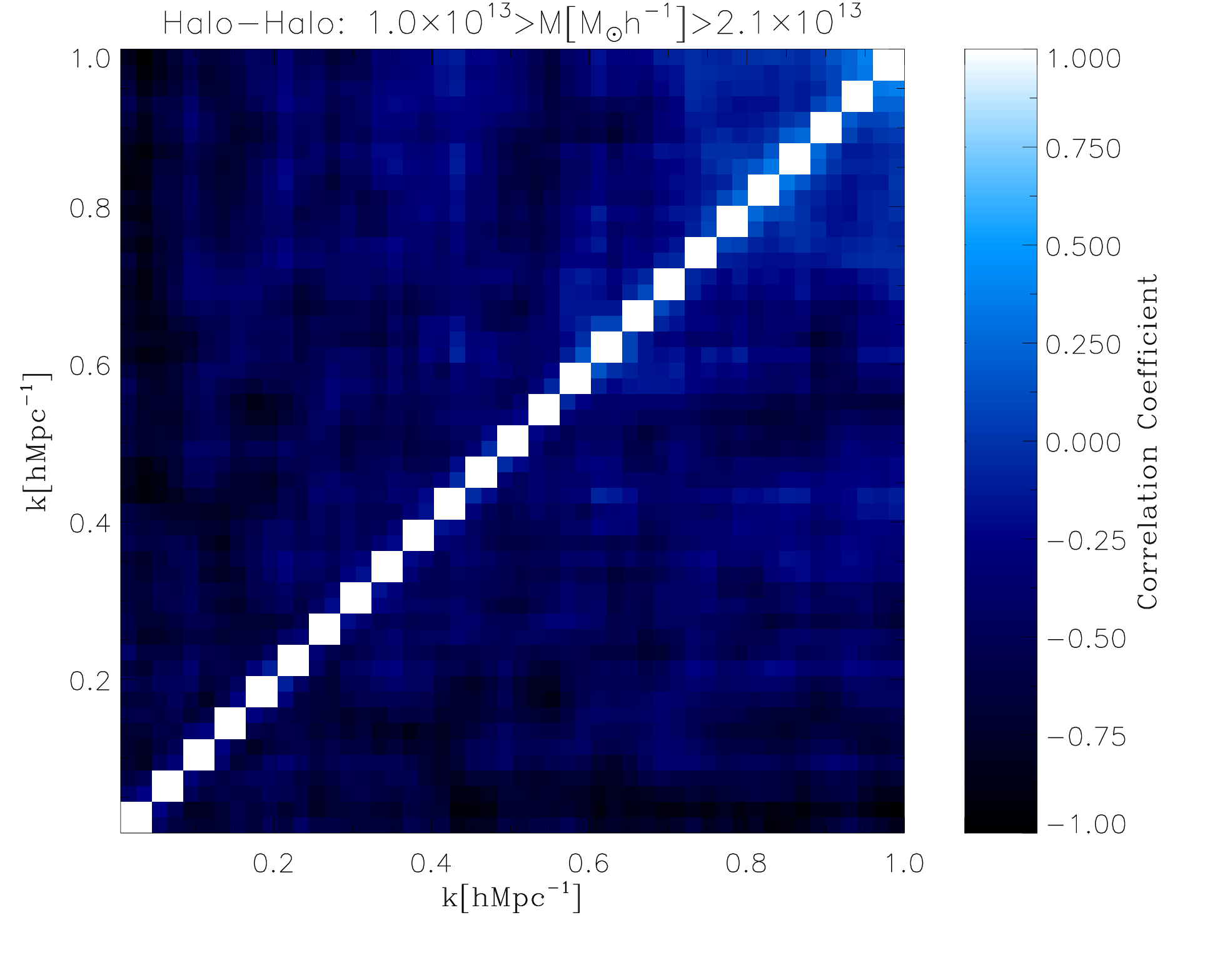}}
\caption{\small{Halo-Halo power spectrum correlation matrices measured
    from the {\tt zHORIZON} simulations.  {\em Top panels:} results
    for power spectra without any correction for shot-noise.  {\em
      Bottom panels:} results after shot-noise correction. {\em Left
      column:} results for the cluster mass halo sample
    ($M>10^{14}\Msol$). {\em Right column:} results for group mass
    haloes ($10^{13}>M[\Msol] >1.38\times10^{13}$).}
\label{fig:CovarianceHH}}
\end{figure*}


\subsection{Results: Mass-mass band-power correlation matrices}

In Fig.~\ref{fig:CovarianceMM} we present the correlation matrices for
the mass-mass power spectrum as measured from the {\tt zHORIZON}
simulations, where the correlation matrices are obtained from,
\be {\mathcal C}[k_i,k_j]=\frac{C[k_i,k_j]}
{\sqrt{C[k_i,k_i]C[k_j,k_j]}}\ .\ee
For the correlation matrices, it was necessary to re-bin the power
spectra. This owed to the fact that when the power is averaged in
shells of thickness the fundamental mode, there are insufficient
numbers of modes on large scales to produce a good $\SN$
\citep{Takahashietal2009}. We therefore chose to re-bin the power by a
factor of 4, and with the contribution from each $k$-shell being
weighted by the number of modes in that shell. Lastly, we box car
smoothed the matrices with a width of two pixels.

In the left panel of Fig.~\ref{fig:CovarianceMM} we show the
correlation matrices obtained from the power spectra without any
shot-noise correction. It can clearly be seen that, going from large-
to small-scales, there is a build up of power correlations between
neighbouring modes and for the smallest scales considered, the matrix
appears perfectly correlated (${\mathcal C}=1$). In the right panel we
show the same, but this time the matrix was generated form the
shot-noise corrected power spectra. There are only small
differences. It is likely that this result owes to the combination of
two facts: firstly, the number density of dark matter particles is
sufficiently high to render the shot-noise contributions to the
covariance of negligible importance
(c.f. Fig.~\ref{fig:CorrMatrixHH}); secondly there is no variation in
the number density of dark matter particles between realizations that
might introduce additional variance (the importance of this will
become clear in the next subsections). Therefore it is likely that the
correlations are purely derived from the the gravitational model
coupling \citep[For a recent and detailed study of the matter power
  spectrum covariance arising due to gravitational instability
  see][]{Takahashietal2009}.


\begin{figure*}
\centerline{ 
\includegraphics[width=8.5cm]{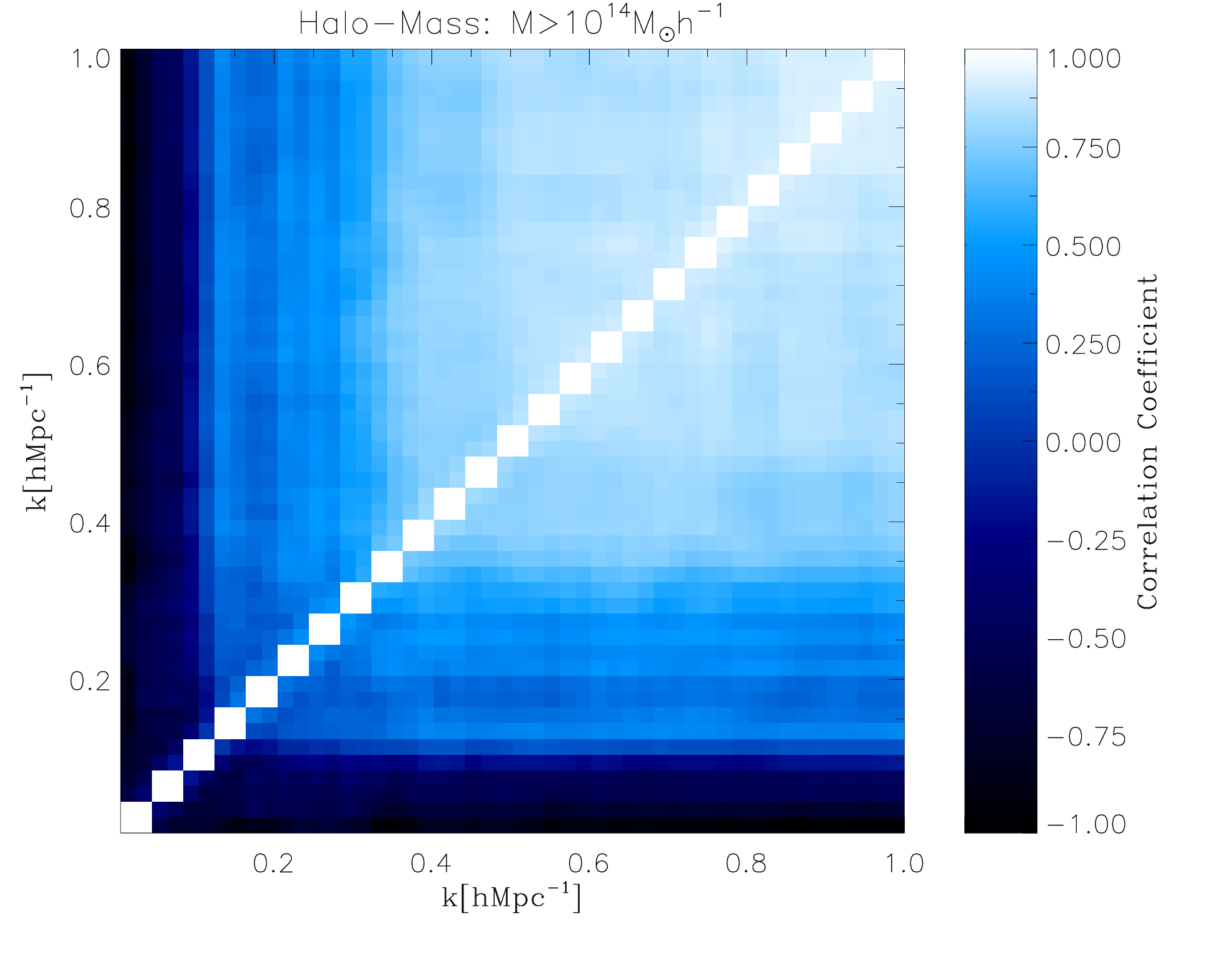}
\includegraphics[width=8.5cm]{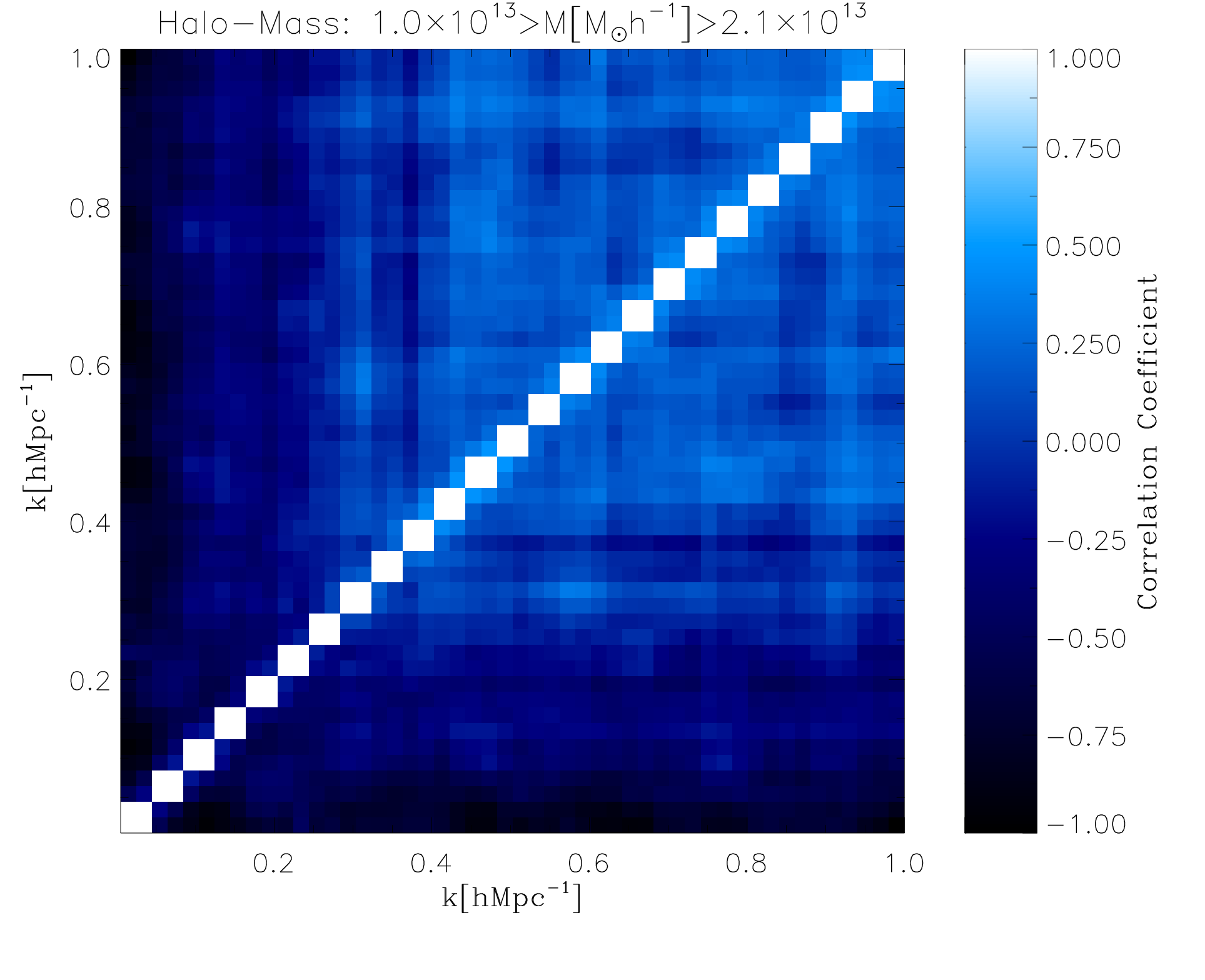}}
\caption{\small{Halo-Mass cross-power spectrum correlation matrices
    measured from the {\tt zHORIZON} simulations.  {\em Left panel:}
    results for a cluster mass halo sample ($M>10^{14}\Msol$). {Right
      panel:} results for a group mass halo sample ($10^{13}>M[\Msol]
    >1.38\times10^{13}$).}
\label{fig:CovarianceHM}}
\end{figure*}


\subsection{Results: Halo-halo band-power correlation matrices}

Fig.~\ref{fig:CovarianceHH} presents the results for the halo-halo
auto-power spectrum correlation matrices. The top two panels show the
results obtained from the power spectra without shot-noise
corrections. The left panel shows results for cluster mass haloes
($M>10^{14}\Msol$) and the right for group mass haloes
($10^{13}>M[\Msol] >1.38\times10^{13}$). We see that the degree of
correlation appears strongly dependent on both the halo mass range
considered and also the scale considered, with the high mass halo
sample having significantly stronger off-diagonal correlations than
for the lower mass sample for a given scale. Both matrices show
significantly more correlation than was found for the dark matter.

In the bottom two panels of Fig.~\ref{fig:CovarianceHH}, we show the
same matrices, but this time constructed from the shot-noise corrected
power spectra. The difference is remarkable -- the strong off-diagonal
correlations that are present in the upper panels has been almost
completely suppressed. The shot-noise corrected covariance matrix may
be written in terms of the shot-noise uncorrected covariance as:
\ba \overline{C}^c_{\h\h}[k_i,k_j] & = &
\left<
\left(\overline{P}^{d}_{\h\h}(k_i)-\frac{1}{\nbar_{\h}}\right)
\left(\overline{P}^{d}_{\h\h}(k_j)-\frac{1}{\nbar_{\h}}\right)
\right> \nonumber \\
& - &
\left<
\overline{P}^{d}_{\h\h}(k_i)-\frac{1}{\nbar_{\h}}
\right>
\left<
\overline{P}^{d}_{\h\h}(k_j)-\frac{1}{\nbar_{\h}}
\right> \ , \\
& = & C^d_{\h\h}[k_i,k_j]- 
{\rm Covar}\left[\frac{1}{\nbar_{\h}},\overline{P}^{d}_{\h\h}(k_i)\right]\nonumber  \\
& - & 
{\rm Covar}\left[\frac{1}{\nbar_{\h}},\overline{P}^{d}_{\h\h}(k_j)\right]+
{\rm Var}\left[\frac{1}{\nbar_{\h}}\right]
\ea
If the number density of the tracer sample does not vary between
realizations, then the shot corrected and un-corrected covariance
matrices are identical. However, if it does then we see that there are
additional sources of covariance that are introduced due to the
coupling between the amplitude of the halo-halo power spectrum and the
mass function of haloes, and from the variance of the number density.
In order for the subtraction of shot-noise to result in a diagonal
correlation matrix, it requires that the cross-correlation between the
halo number counts and the halo power spectrum cancel with the
off-diagonal contributions to $C^d_{\h\h}[k_i,k_j]$. It is beyond the
scope of this current work to illuminate this issue further and it
shall remain as a topic for future investigation. One caveat to the
above results, is that it is well known that the standard shot-noise
correction is too strong for haloes, since it results in negative
power on small scales \citep{Smithetal2007}. It is therefore likely
that this will have some impact on the covariance matrix.

Lastly, we now see that for the matter power spectra, since the number
density of dark matter particles does not vary between realizations,
then we must have $\overline{C}^c_{\delta\delta}[k_i,k_j] =
\overline{C}^d_{\delta\delta}[k_i,k_j]$.


\subsection{Results: Halo-mass band-power correlation matrices}

In Fig.~\ref{fig:CovarianceHM} we show the correlation matrices for
the halo-mass cross-power spectra. The left panel shows results for
cluster mass haloes ($M>10^{14}\Msol$) and the right for group mass
haloes ($10^{13}>M[\Msol] >1.38\times10^{13}$). Similar to the halo
auto-power correlation matrix, we see that the degree of correlation
appears strongly dependent on both halo mass and scale. Interestingly,
we note that whilst the spectra from the high-mass sample show more
band-power correlation than for the dark matter, the lower mass halo
sample appears to show less. This further recommends the cross-spectra
approach for further investigation as an improved estimator for
large-scale structure.

All of the above matrices serve to warn us that, whilst the Gaussian
plus Poisson model describes the diagonal errors reasonably well, it
fails to capture the build up of correlations between Fourier
modes. To describe the above results one must model both the full
non-Gaussian trispectrum generated by gravitational mode-coupling
\citep{Scoccimarroetal1999b,Takahashietal2009} and, as we have shown
in this paper, the covariance introduced by the point sampling for the
mass tracers.


\begin{figure*}
\centerline{ 
\includegraphics[width=13cm]{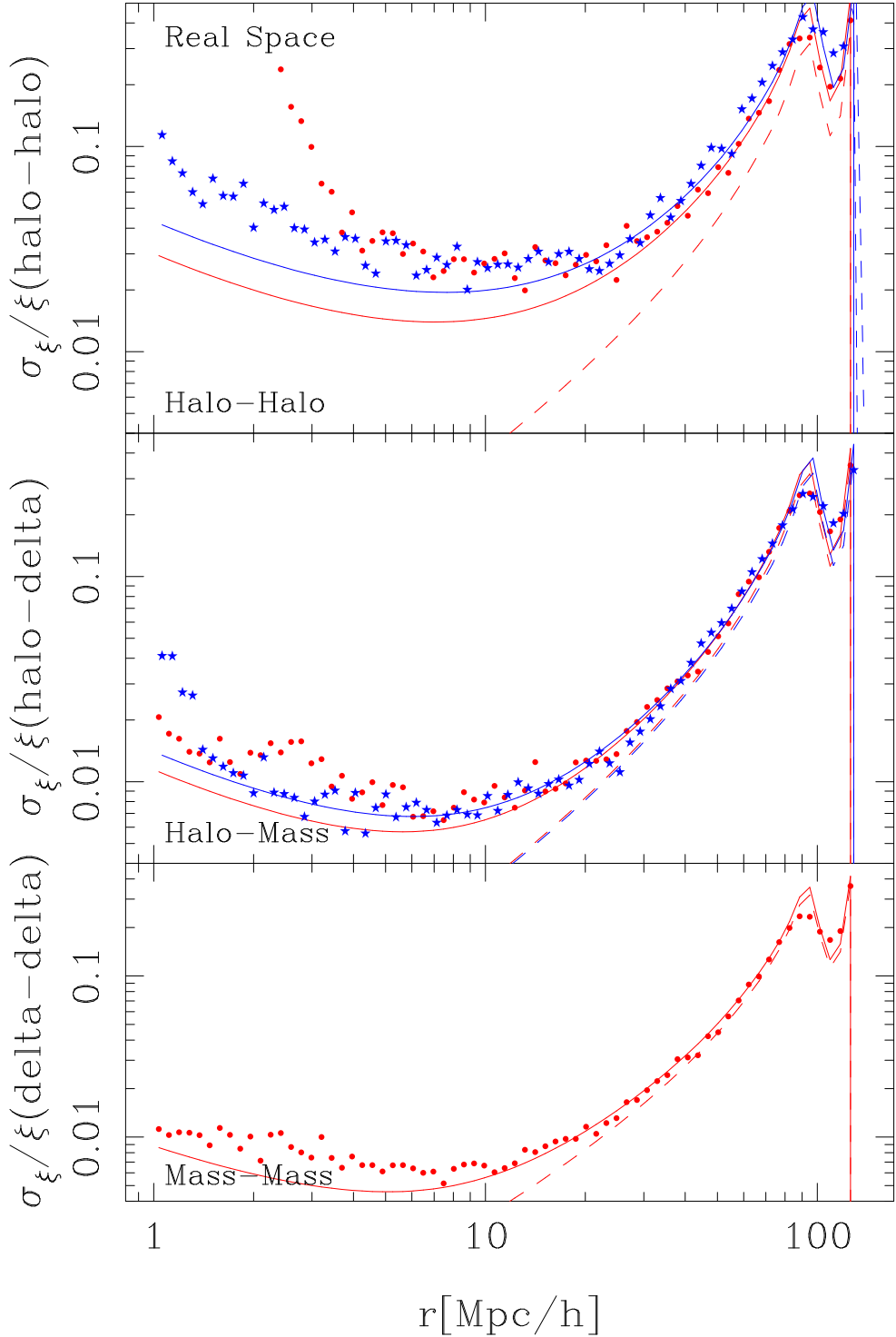}}
\caption{\small{Comparison of the fractional variance of mass and halo
correlation functions as measured from the {\tt zHORIZON}
simulations with theoretical predictions.  Similar to
Fig.~\ref{fig:PowError}, the three panels show the standard deviation
in the bin averaged correlation functions, ratioed to the mean
correlation function, as a function of the spatial scale. From top to
bottom the panels show results for the halo-halo, halo-mass and
mass-mass correlations. Symbols show estimates measured from the
$N$-body simulations. In the top two panels, the (red) point symbols
and the (blue) star symbols denote haloes with masses in the range
$(M>1.0\times 10^{14} \Msol)$ and
$(1.0\times10^{13}<M[\Msol]<2\times10^{13})$, respectively.  Again, the
solid lines represent the theoretical predictions from the Gaussian
plus Poisson sampling theory. Dash lines represent the pure Gaussian
predictions.}
\label{fig:XiError}}
\end{figure*}


\subsection{Results: band-correlation function variances}

As a final study we now consider the correlation function errors.  The
main advantage of the configuration space is that the constant
shot-noise correction, which is necessary for the power spectra are
not required here. This follows from the fact that the Fourier
transform of a constant gives a delta function at zero lag. However,
as was described in \S\ref{sec:CovCorr}, the shot-noise corrections do
affect the correlation function errors.

In Figure~\ref{fig:XiError} we present measurements from the ensemble
of {\tt zHORIZON} simulations for the fractional errors on the
mass-mass (bottom panel), the halo-mass (middle panel) and the
halo-halo (top panel) correlation functions. Again, we only show
results for the highest and lowest bins in halo mass. The correlation
functions were generated using the {\tt DualTreeTwoPoint} code, which
is a parallel, tree-based algorithm and is described more fully in
Smith et al. (in preparation). For the dark matter sample, we used
roughly $\sim4\times10^6$ particles, sub-sampled from the available
$\sim4\times10^8$ for each estimate.

The main result to note from this analysis is that, whilst for the
power spectrum on large scales the fractional error is the same
irrespective of tracer, this is not the case for the correlation
function. We note that for $r>20\Mpc$, the halo-mass cross-correlation
appears to be a more efficient estimator than the simple
auto-correlation function, by almost a factor of $\sim2$. To make this
statement more concrete we should include the off-diagonal errors in
the calculation of $\SN$. However, from our discussion in
\S\ref{sec:CovCorr}, we expect that the off-diagonal errors are also
reduced.  We shall reserve this for a future work.

Another important point to note, is that in nearly all cases the
theoretical predictions for the Gaussian plus Poisson sampling error
estimates are an underestimate of the measured errors, especially on
scales $(r<20\Mpc)$. The predictions being worst for the
auto-correlation function for the high mass halo sample, and this is
in agreement with the power spectrum results from the previous
section. 

The errors in the auto-correlation functions were previously
investigated in numerical simulations by \cite{Smithetal2008a} and
\citet{Sanchezetal2008}, who showed that the Gaussian plus Poisson
model provided a good description at the scale of the Baryonic
Acoustic Oscillations ($r\sim100\Mpc$). Our results extend this
analysis to the cross-correlation functions.  Also, the range of
investigated scales is extended to smaller scales by more than one
order of magnitude.


\section{Conclusions}\label{sec:conclusions}

In this paper we have performed a detailed investigation of the errors
associated with auto- and cross-power spectra and also the
cross-correlation function of different tracers of the density field.

In \S\ref{sec:background} we developed the counts-in-cells approach
for a multi-tracer approach to the clustering statistics. We showed
that not all cross-power spectra are free from a shot-noise
correction, and that the precise correction one should apply depends
on the sampling distribution function.

In \S\ref{sec:crosscovar} we gave a derivation of the full
non-Gaussian covariance matrix for the cross-power spectrum, including
all the sources of variance that arise from the Poisson sampling of
the mass tracers and this extends the standard results
\citep{Scoccimarroetal1999b,MeiksinWhite1999,Cohn2006,Hamiltonetal2006}.
We showed that, for the case of Poisson sampling of Gaussian
fluctuations, there were terms that contributed to the off-diagonal
terms of the covariance matrix. We showed that in the small-scale
limit $k\rightarrow\infty$ these terms dominate over all other sources
of variance (including Non-Gaussian terms generated from gravitational
mode-coupling) and the covariance matrix becomes perfectly correlated.

In \S\ref{sec:Efficiency} we investigated the efficiency of the
cross-power spectrum. We used the relative signal-to-noise $(\SN)$
ratio of two different estimators as a diagnostic for efficiency. For
the case where a high-density sample of tracers was cross-correlated
with a low-density sample, it was shown that the former approach was a
more efficient estimator than the case where one simply
auto-correlates the low-density sample. As an example, we showed that
for the determination of cluster bias, the cross-power spectrum
approach would yield significant gains in $\SN$.  Other uses are
improving estimates of the luminosity dependence of galaxy bias.

In \S\ref{sec:CovCorr} we explored the covariance of auto- and
cross-correlation functions. It was shown that whilst the correlation
function covariance matrix in general is not diagonal, the
discreteness terms that led to off-diagonal covariance in the power
spectrum do not generate off-diagonal elements in the correlation
function covariance. Thus the correlation function covariance matrix
appears easier to understand and model than the power spectrum
covariance.

In \S\ref{sec:simulations} we used a large ensemble of $N$-body
simulations, to obtain estimates of the power spectrum and correlation
function errors.  We showed for the fractional errors on the mass-mass
halo-mass and halo-halo spectra, that the numerical results were in
reasonably good agreement with the Gaussian plus Poisson sampling
model, but the measurements showed larger variance than the theory. It
was also shown that in the limit of large scales and in the case that
Poisson error is not dominant, then the fractional errors for all
spectra are equivalent, since they are simply $\propto
k^{-1}\Vu^{-1/2}$.  

We investigated the correlation matrix for the mass-mass power
spectrum, and confirmed that there were strong correlations between
different band-powers
\citep{Scoccimarroetal1999b,MeiksinWhite1999,Takahashietal2009}.  We
showed that correcting the spectra for shot-noise does not change
the correlation matrix significantly. We investigated the halo-halo
auto-power covariance matrix without applying a correction for
shot-noise. We showed that the degree of correlation increased with
the mass of the halo sample considered and that the matrices showed
more band-power correlation than for the dark matter. We then
estimated the covariance from the shot-noise corrected spectra and
found that the off-diagonal errors were dramatically reduced, almost
decorrelating individual band-powers.  We conjectured that this arises
from the subtraction of the covariance between halo number density and
the halo-halo power spectrum and also the variance in the halo number
density. We investigated the cross-power correlation matrix for
haloes and dark matter and showed that the correlations were reduced
compared to the shot-noise uncorrected halo-halo matrices and for the
lowest-mass halo sample, were less correlated than the dark matter.
We investigated the errors in configuration space, and showed
that there was a significant gain in $\SN$ on all scales from using
the cross-correlation function of haloes and dark matter as opposed to
simply examining the auto-correlation function of haloes.

We conclude that, for certain cases, the cross-spectra and
cross-correlation functions are more efficient probes for the
large-scale structure, than the standard auto-spectra and
auto-correlation function approaches that are widely in use. These
cases concern studies aiming to measure: the luminosity dependence of
the galaxy bias \citep{Norbergetal2002a,Tegmarketal2004b}; the cluster
bias as a function of mass and hence constrain the degree of
primordial non-Gaussianity in the initial conditions
\citep{Dalaletal2008,Slosar2008,Desjacquesetal2009,Pillepichetal2008}.

\vspace{0.2cm}


\section*{Acknowledgements}

RES acknowledges: R.~Angulo, V.~Desjacques, C.~Porciani,
R.~Scoccimarro, U.~Seljak, and R.~Sheth for useful discussions;
A.~Saintonge for help with IDL. RES kindly thanks L. Marian for
comments on the draft. RES thanks V.~Springel for making public his
{\tt GADGET-2} code and for providing his {\tt B-FoF} halo finder,
R.~Scoccimarro for making public his {\tt 2LPT} initial conditions
code and U.~Seljak and M.~Zaldarriaga for making public their {\tt
  cmbfast} code. RES acknowledges support from a Marie Curie
Reintegration Grant, and the Swiss National Foundation.  When this
paper was in the refereeing stage, two related works appeared on the
arXiv: \citet{HutsiLahav2008} and \citet{Whiteetal2008}.



\bibliographystyle{mn}


\end{document}